\documentclass[twocolumn,10pt]{article}

\usepackage[a4paper,margin=1.7cm,columnsep=0.65cm]{geometry}
\usepackage{xcolor}
\usepackage{graphicx}
\usepackage{amsmath,amssymb,siunitx}
\usepackage{bm}
\usepackage{float}
\usepackage{titlesec}
\usepackage{hyperref}
\usepackage{subfiles}
\usepackage{svg}
\usepackage{graphicx}
\usepackage{cuted}
\usepackage{caption}
\usepackage{authblk}

\makeatletter
\renewcommand{\AB@authnote}[1]{}
\makeatother

\date{}
\usepackage[
    backend=biber,
    style=nature,
    sorting=none,
    doi=false,
    url=false,
    isbn=false,
    eprint=false
]{biblatex}

\newif\ifwithsupplementary
\withsupplementarytrue

\graphicspath{{./}{Figures/}{Supplementary_Figures/}}

\titleformat{\section}
  {\normalfont\bfseries\large\raggedright}
  {}
  {0pt}
  {\MakeUppercase}

\titlespacing*{\section}
  {0pt}
  {1.2em}
  {0.45em}

\titleformat{\subsection}
  {\normalfont\bfseries\normalsize\raggedright}
  {}
  {0pt}
  {}

\titlespacing*{\subsection}
  {0pt}
  {1.0em}
  {0.35em}
\DeclareUnicodeCharacter{2212}{\ensuremath{-}}
\title{\textbf{Cavity Control of Emission Dynamics from Spin Defects in Hexagonal Boron Nitride}}
\author{%
{Anuj Kumar Singh\textsuperscript{1,4,*,$\dagger$}},
{Atanu Patra\textsuperscript{1,4,*,$\ddagger$}},
{Paul Konrad\textsuperscript{2}},
{Subhamoy Sahoo\textsuperscript{1,4}},
{Ankur Arora\textsuperscript{1,4}},
{Johannes D\"ureth\textsuperscript{1,4}},
{Georgy V. Astakhov\textsuperscript{3}},
{Monika Emmerling\textsuperscript{1,4}},
{Martin Kamp\textsuperscript{1,4}},
{Andreas Sperlich\textsuperscript{2,4}},
{Vladimir Dyakonov\textsuperscript{2,4}},
{Sven H\"ofling\textsuperscript{1,4}},
{Simon Betzold\textsuperscript{1,4,$\S$}}}

\affil[1]{Julius-Maximilians-Universität Würzburg, Physikalisches Institut, Lehrstuhl für Technische Physik, Am Hubland, Würzburg, Deutschland}
\affil[2]{Julius-Maximilians-Universität Würzburg, Physikalisches Institut, Lehrstuhl für Experimental Physics 6, Am Hubland, Würzburg, Deutschland}
\affil[3]{Helmholtz-Zentrum Dresden-Rossendorf, Institute of Ion Beam Physics and Materials Research, Dresden, Germany}
\affil[4]{Julius-Maximilians-Universität Würzburg, Physikalisches Institut and Würzburg-Dresden Cluster of Excellence ctd.qmat, Am Hubland, Würzburg, Germany}

\begin{document}

% Main-paper bibliography begins here
\begin{refsection}[ref.bib]
\maketitle

\begingroup
\renewcommand{\thefootnote}{\fnsymbol{footnote}}

\footnotetext[1]{These authors contributed equally to this work.}
\footnotetext[2]{
 \texttt{anuj-kumar.singh@uni-wuerzburg.de}}
\footnotetext[3] {\texttt{atanu.patra@uni-wuerzburg.de}}
\footnotetext[4] {\texttt{simon.betzold@uni-wuerzburg.de}}.

\endgroup
\begin{abstract}
Optically addressable negatively charged boron vacancy ($\mathrm{V_B^-}$) defects in hexagonal boron nitride (hBN) are promising for quantum sensing at room temperature. The two-dimensional structure of hBN allows direct access to the local environment, enabling proximal sensing of target materials. However, its broadband emission and low quantum yield hinder efficient light–matter interaction, and the coupling of $\mathrm{V_B^-}$ optical dipoles to confined cavity fields remains poorly understood. In this study, we integrate  $\mathrm{V_B^-}$ defects with a tunable plano-concave microcavity, achieving three-dimensional optical confinement. This approach yields photoluminescence (PL) enhancement and cavity-modified optical decay dynamics, providing insights into the orientation of the $\mathrm{V_B^-}$ defect dipole. Time-resolved PL measurements identify a distinct fast decay component, suggesting a complex dipole geometry for the defects. Therefore, a tilted optical dipole model is considered to attribute the fast and slow decay components to in-plane and out-of-plane dipole contributions, respectively. Additionally, optically detected magnetic resonance and excitation-polarisation-dependent PL measurements provide further insights into local symmetry breaking and potential optical dipole configurations. These results highlight the role of optical dipole orientation in $\mathrm{V_B^-}$ defect–cavity coupling.

\end{abstract}


\maketitle

\section{Introduction}

Quantum sensing provides access to weak physical signals with high sensitivity and spatial resolution by exploiting the response of a well-controlled quantum system to its local environment  \cite{degen_quantum_2017}. A broad range of physical platforms has been developed for quantum sensing, including neutral atoms, Rydberg atoms, superconducting circuits, and solid-state spin systems \cite{degen_quantum_2017}. Among these, solid-state spin defects are particularly attractive for nanoscale sensing because their spin states can be initialised, manipulated, and optically read out, enabling local measurements of magnetic and electric fields, temperature, pressure, and strain \cite{Gottscholl2021}.
\\
One of the most established platforms for solid-state quantum sensing is the negatively charged nitrogen-vacancy (NV) centre in diamond, owing to its combination of room-temperature operation, optical spin initialisation and readout, and long spin coherence times \cite{Doherty.2013, casola_probing_2018}. However, the three-dimensional nature of diamond  as well as other 3D platforms imposes practical constraints on how closely the active spin can be positioned to a target, particularly while maintaining stable optical and spin properties near the surface. 
Two-dimensional (2D) van der Waals (vdW) materials have recently emerged as promising platforms for quantum photonic technologies, such as quantum sensing, addressing these challenges \cite{Wang2012,Turunen2022,Stern2024,Kumar,Piccinini2025}. Among them, hBN provides a versatile solid-state host for defect states that can behave as artificial atoms and serve as building blocks for quantum photonic architectures \cite{Kianinia2022,Tran2015,Singh2025,Cai2024,PalaciosBerraquero2016}.
Besides several defects in hBN that act as single-photon emitters \cite{Kumar2024,Hazra2026}, the $\mathrm{V_B^-}$ remains the only known defect with a clearly elucidated structure. This is especially interesting because it hosts a triplet system that exhibits a spin-dependent optical response \cite{Gottscholl2020}. The $\mathrm{V_B^-}$ centre has a spin triplet ground state (GS) with an axial zero-field splitting (ZFS), D $\approx$ 3.5 GHz and a non-axial ZFS, E $\approx$ 50 MHz \cite{Mathur2022}. The spin states can be optically polarised under visible excitation with an optimal excitation energy of ~2.6 eV \cite{Baber.2022}. The 2D nature of hBN also allows the defects to be situated in close proximity to the surface, which is an advantage for quantum sensing applications \cite{Gao2021, Durand.2023, Moon2024}.
Despite these advantages, the optical performance of $\mathrm{V_B^-}$ centres remains limited by their low quantum efficiency of 0.03\% \cite{Reimers2020}. This limitation mainly arises from dominant non-radiative intersystem crossing (ISC) processes, which compete strongly with radiative recombination \cite{Reimers2020}.
\\
In this context, optical microcavities offer a promising route to improve weak optical emission by modifying the local density of optical states and enhancing radiative emission through the Purcell effect \cite{PhysRev.69.674.2}. Cavity-enhanced emission from  V$_\mathrm{B}^-$ centres has already been demonstrated using different  plasmonic architectures and monolithic bullseye cavities \cite{mendelson_coupling_2022, qian_unveiling_2022, xu_greatly_2023, moon_fiber-integrated_2024,Frch2021}. These studies show enhanced PL and, in some cases, improved optically detected magnetic resonance (ODMR) readout, establishing cavity integration as a promising approach to improving the optical response of  V$_\mathrm{B}^-$ spin defects.
 \begin{figure*}[!t]
    \centering
     \includegraphics[width=\textwidth]{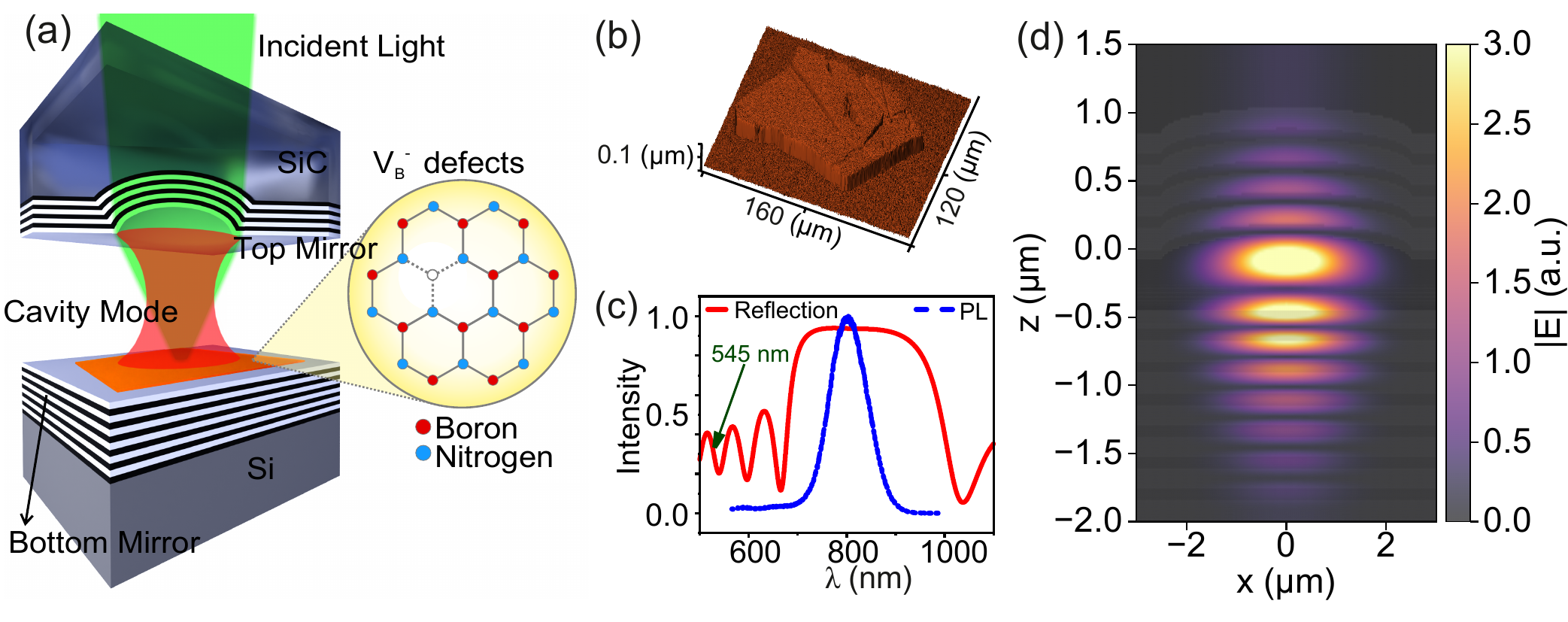}
      \captionof{figure}{\textbf{hBN defect centres coupled to a tunable open microcavity.} \textbf{(a)} Schematic illustration of the tunable open microcavity used to couple emission from $\mathrm{V_B^-}$ centres in an hBN flake. \textbf{(b)} Laser scanning microscopy image of the selected hBN flake, which has a thickness of approximately 100 nm and is used for the subsequent optical measurements. \textbf{(c)} Calculated reflectivity spectrum of the bottom DBR  obtained using the transfer-matrix method. The stop band exhibits a reflectivity of approximately 90\% over the spectral region relevant to the defect emission, while optical excitation is performed at 545~nm. The blue curve indicates that the broadband emission from the $\mathrm{V_B^-}$ defects lies within the DBR stop band, close to its central high-reflectivity region. \textbf{(d)} Simulated electric-field distribution obtained using Lumerical FDTD for a dipole positioned on the hBN surface inside the full cavity.}
    \label{cavity}
\end{figure*}
However, these approaches are predominantly based on fixed cavity geometries, where the resonance wavelength, mode volume, and spatial overlap with the emitter are largely determined during fabrication. A tunable open microcavity offers a complementary approach, where the cavity resonance and mode volume can be controlled through cavity length, allowing systematic optimisation of the emitter–cavity interaction. This platform is particularly relevant for poor emitters, where reducing the cavity mode volume and funnelling emission into a well-defined optical mode can enhance radiative efficiency beyond simple spectral filtering. For V$_\mathrm{B}^{-}$, another important consideration is the orientation of the optical transition dipole relative to the cavity electric field, since this directly determines the emitter–cavity coupling strength and the achievable Purcell enhancement. Theoretical studies predict strongly polarisation-dependent optical transitions for V$_\mathrm{B}^{-}$ \cite{Ivdy2020}. However, a direct experimental determination of the V$_\mathrm{B}^{-}$ emission-dipole orientation is still lacking  \cite{Sortino2024}, making it an important parameter for the quantitative design of V$_\mathrm{B}^{-}$ cavity interfaces.
\\
In this work, $\mathrm{V_B^-}$ centres in hBN are integrated into a tunable open microcavity to investigate light–matter coupling and emission dynamics. In this configuration, the hBN flake is positioned between two distributed Bragg reflector (DBR) mirrors, enabling tuning of the optical mode by adjusting the cavity length. This tunability permits the cavity resonance to overlap spectrally with the broadband defect emission, thereby providing an active method for controlling the emitter–cavity interaction.
Insights into the optical dipole characteristics of $\mathrm{V_B^-}$ defects are obtained by resolving a distinct fast decay component in time-resolved photoluminescence (TRPL) measurements. The Purcell-enhanced modification of this component, supported by the in-plane dipole simulation in the tunable cavity supports the hypothesis of a tilted optical dipole exhibiting varying coupling strengths to the confined cavity field. Consequently, the local symmetry and polarisation-dependent optical response of the defects are investigated using ODMR and excitation-polarisation-dependent PL spectroscopy to elucidate the possible origin of these dipole contributions and their role in the observed cavity enhancement.

\section{Results and discussion}
\subsection{Cavity Design}

The tunable open microcavity is designed to couple the broadband emission from $\mathrm{V_B^-}$ centres to a confined optical mode, as schematically shown in Fig.~\ref{cavity}(a). The cavity is formed between a planar bottom DBR on silicon (Si) and a locally curved top lens fabricated on a transparent silicon carbide (SiC) substrate. Concave lenses of various diameters are fabricated by focused-ion-beam milling and subsequently coated with the top DBR (Supplementary Fig.~\ref{cavity_align}). The local curvature provides transverse optical confinement and reduces the cavity mode volume, while the open geometry allows continuous tuning of the cavity resonance. The substrate selection for the bottom and top mirrors is discussed in Supplementary  Section~\ref{sec:substrate} and Fig.~\ref{substrate}.

The hBN flake was uniformly irradiated with 30 keV nitrogen ions to generate to achieve a homogeneous distribution of optically active $\mathrm{V_B^-}$ spin defects. We then transferred this flake onto the bottom DBR. Laser-scanning microscopy of the selected flake Fig.~\ref{cavity}(b)) shows its morphology and lateral extent, as well as the approximate thickness of 100 nm. Atomic-force microscopy confirms a thickness of $\sim$ 100 nm (Supplementary Fig.~\ref{AFM}), while PL mapping confirms spatially uniform defect emission over the selected region (Supplementary Fig.~\ref{Raman-Pl}). 
Furthermore, the $\mathrm{V_B^-}$ density was estimated to be approximately $10^{16}$~defects~cm$^{-3}$ using the method described in Ref. \cite{Patra2025}, as detailed in Supplementary Section~\ref{sec:RPL}.

The DBR design was done by transfer-matrix calculations, shown in Fig.~\ref{cavity}(c). The resulting stop band reaches approximately 90\% reflectivity across the broadband $\mathrm{V_B^-}$ emission, while the 545 nm excitation wavelength lies near a reflectivity minimum. The measured defect emission overlaps with the central region of the stop band, allowing the cavity resonance to be tuned through the emission band. Details of the number of DBR layer pairs and their thicknesses are provided in Supplementary Section~\ref{sec:cavityalign}. The finite-difference time-domain (FDTD)-simulated cavity-mode profile is shown in Fig.~\ref{cavity}(d). The electric field distribution of the dipole is confined along the cavity axis, with a longitudinal electric-field antinode positioned close to the hBN interface. This provides favourable spatial overlap between the $\mathrm{V_B^-}$ defects and the resonant cavity field.

\subsection{Cavity-coupled emission}
\begin{figure*}[t]
\centering
    \includegraphics[width=1.0\textwidth]{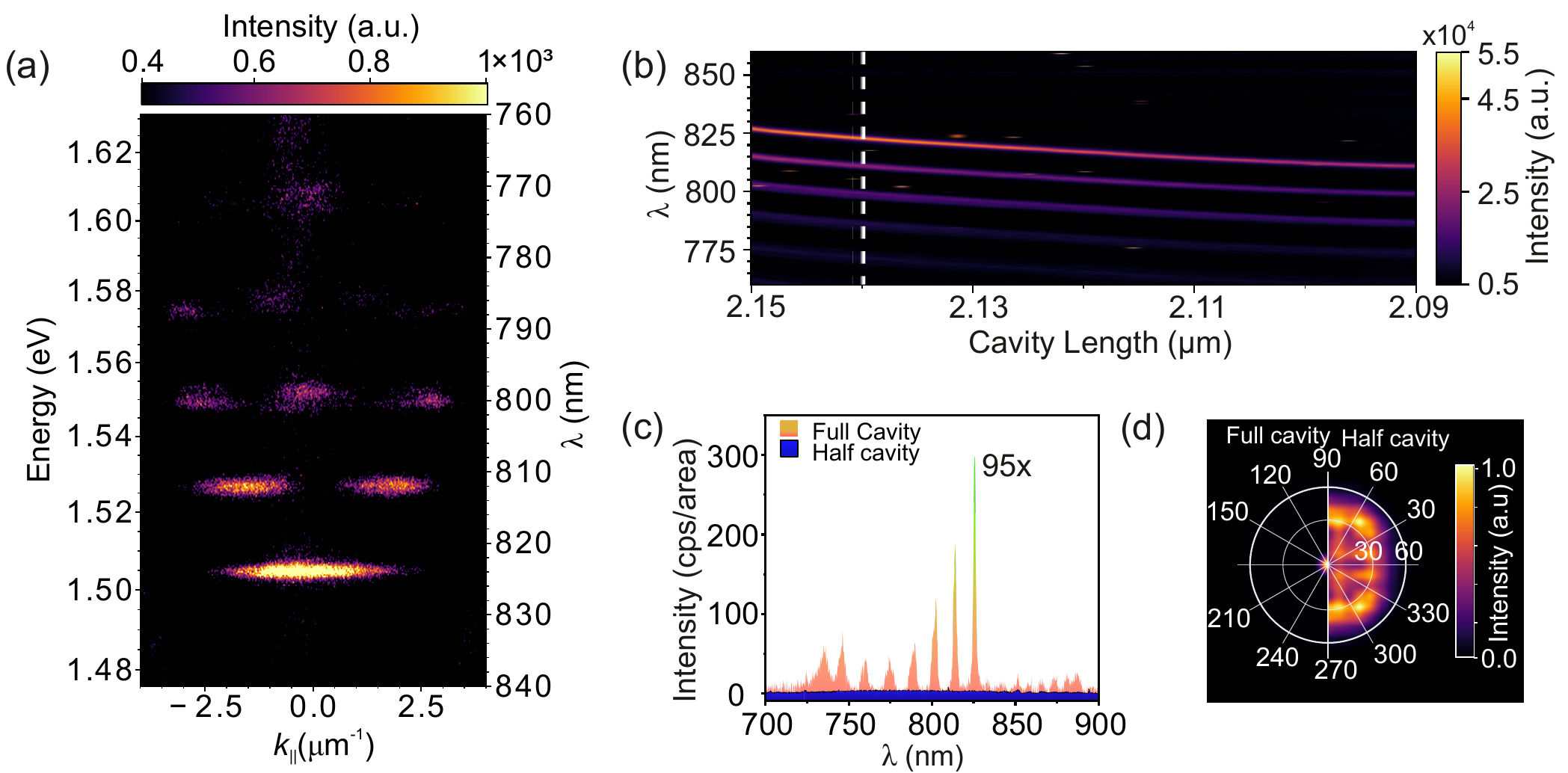}
 \captionof{figure}{\textbf{PL from defect emitters coupled to a tunable open microcavity.}
\textbf{(a)} Angle-resolved PL from the tunable open microcavity for a concave mirror with a lens diameter of 5~\textmu m and air gap of 725 nm. The emission pattern exhibits well-defined Laguerre-Gaussian-like  transverse cavity modes, including the fundamental and higher-order modes (such as $s$, $p$, and $d$-like modes), indicating efficient optical confinement and coupling of the defect emission to the cavity. \textbf{(b)} Evolution of the cavity mode as a function of cavity length, demonstrating spectral tuning of the resonant mode wavelength. As the cavity length is varied, the cavity resonance is tuned across the defect emission band. A pronounced enhancement of the fundamental mode is observed when the emitter is brought into spectral resonance with the cavity, evidencing cavity-enhanced light extraction and emission. \textbf{(c)} PL spectrum of the defect emission in the uncoupled configuration (blue) and in the cavity-coupled configuration (red-to-green gradient fill), measured from the same flake region.  \textbf{(d)} The simulated far-field emission distributions of the dipole are compared for the half- and full-cavity configurations. In the full cavity, the emission exhibits a more confined angular distribution, indicating enhanced directionality.}
    \label{Pl}
\end{figure*}
The angle-resolved PL spectrum at a fixed cavity length is shown in Fig.~\ref{Pl}(a). The measurement was performed using the 5~$\mu$m-diameter concave structure, with a depth of 185~nm and an estimated radius of curvature of approximately 17~$\mu$m. A comparison with 4, 5 and 6~$\mu$m structures is provided in Supplementary Figs.~\ref{D4}, \ref{D6}, and \ref{PL raw}. 

A well-defined ladder of Laguerre–Gaussian-like \cite{Betzold2017} transverse modes is observed, extending from the fundamental mode to higher-order modes up to $m=5$. These discrete resonances arise from transverse quantisation of the optical field in the stable plano-concave resonator. The concave lens provides transverse confinement, while the two DBRs define the longitudinal cavity resonance \cite{Benedikter2015}. Starting from the bottom of the observed mode ladder, the resonances can be classified by their transverse order as s-like ($m=0$), p-like ($m=1$), d-like ($m=2$), and higher-order modes.  The transverse-mode spacing is determined by the Gouy phase and therefore depends on both the cavity length and the radius of curvature of the concave mirror  \cite{Kogelnik1966}. The resolved mode ladder confirms transverse confinement of the cavity field and coupling of the broadband defect emission into the cavity modes. The details of the measurement procedure are  shown in Supplementary Section~\ref{sec:arm}, and Supplementary Fig.~\ref{arm}. Details of the cavity alignment procedure are given in Supplementary Section~\ref{sec:cavityalign} and Supplementary Fig.~\ref{cavity_align}.

We next tuned the mirror separation to sweep the cavity resonance across the $\mathrm{V_B^-}$ emission band. A contour plot of the PL spectra as a function of cavity length is shown in Fig.~\ref{Pl}(b). The separation between mirrors was controlled by the voltage applied to the piezo positioning stage. As the mirror separation changes, the longitudinal cavity resonance shifts continuously through the defect emission. The integrated cavity PL reaches a pronounced maximum at an effective cavity length of approximately 2.14~$\mu$m (which corresponds to the 725 nm air gap between the top and bottom mirror; see  Supplementary Section~\ref{sec:ag} and Supplementary Fig.~\ref{air-gap} ), where the fundamental cavity mode has the strongest spectral overlap with the defect emission. The enhancement around the white dotted line in Fig.~\ref{Pl}(b) extends over a cavity-length range of approximately 5~nm, reflecting the finite spectral width of the cavity resonance together with the broadband $\mathrm{V_B^-}$ emission.
To understand the spectral intensity enhancement in more detail, we compare the PL spectra measured under resonant cavity conditions with those obtained in the half-cavity reference configuration, as shown in Fig.~\ref{Pl}(c). When the $\mathrm{V_B^-}$ emission is spectrally coupled to the cavity mode, a pronounced increase in the collected PL intensity is observed. For the fundamental cavity mode, the directly measured PL intensity increases by up to a factor of 20, as shown in Supplementary Fig.~\ref{PL raw}. The transmission values of the dipole emitter in the FDTD simulation also support these PL data, as shown in Supplementary Fig.~\ref{T-FDTD}.  After accounting for the area normalisation for the half and full-cavity configurations (see Supplementary Section~\ref{sec:NF}), the enhancement reaches approximately 95-fold in Fig.~\ref{Pl}(c). This enhancement can arise from the combined effects of emission redistribution into the resonant cavity mode, improved emission directionality and modification of the local density of optical states. We therefore examine these contributions separately in the following sections.

\begin{figure*}[t]
\centering
     \includegraphics[width=\textwidth]{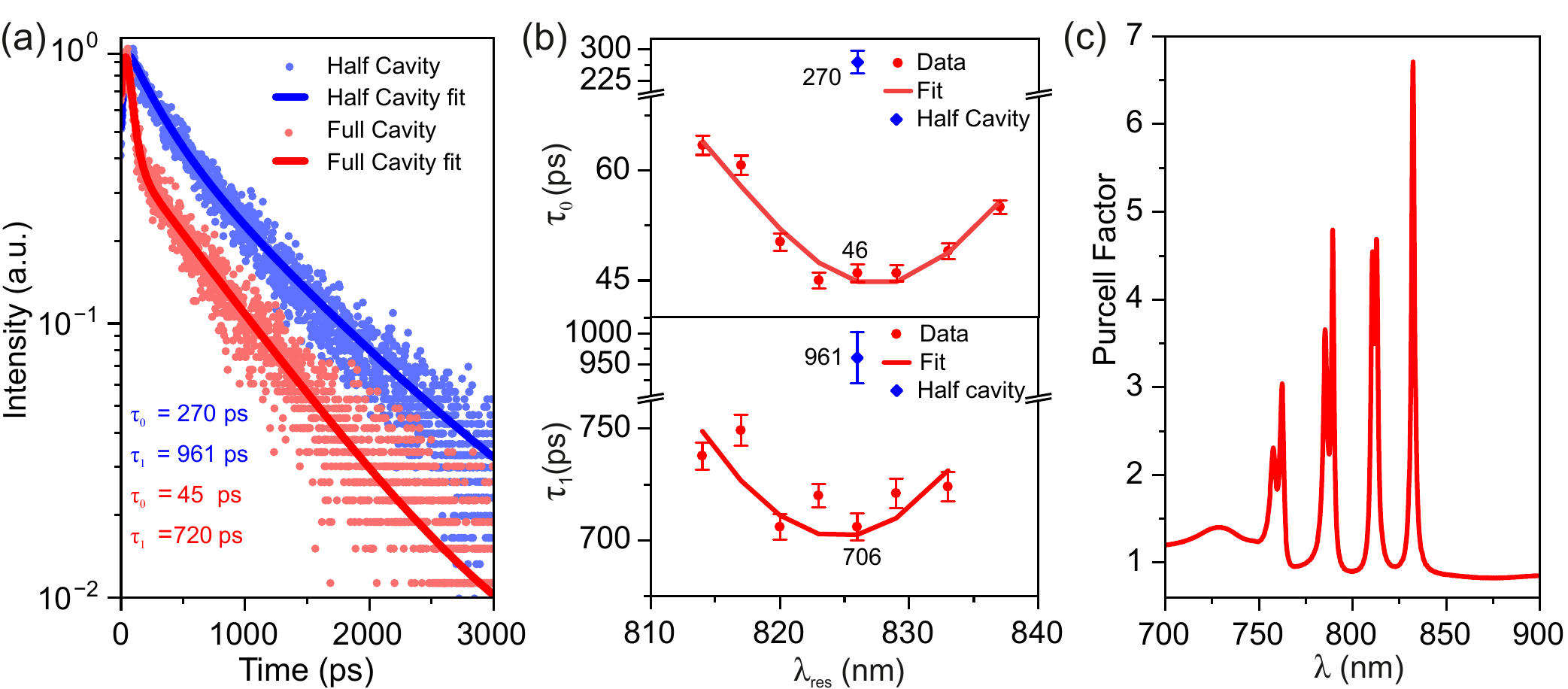}
    
\captionof{figure}{\textbf{Lifetime modification and Purcell enhancement of hBN defect emission in a tunable open microcavity.}
\textbf{(a)} TRPL decay measured for the cavity-coupled emission from $\mathrm{V_B^-}$ centres under pulsed excitation at 545~nm. The red curve represents the decay measured when the defect emission is coupled to the cavity mode, while the blue curve shows the corresponding reference decay measured in the half-cavity configuration. Both decay traces are fitted using a biexponential decay model, allowing comparison of the modified emission dynamics induced by the cavity environment. \textbf{(b)} Evolution of the PL decay as the fundamental cavity mode is tuned into and out of resonance with the defect emission. A reduction in the measured lifetime is observed near the resonance condition, indicating an enhanced radiative decay rate due to the Purcell effect. The detailed fitting procedure is described in the in the Supplementary section S3. \textbf{(c)} Simulated Purcell factor for the in-plane dipole component, showing the enhancement of the emission rate inside the full-cavity configuration. }
    \label{Lifetime}
\end{figure*}

\textbf{\textit{Angular Distribution:}} We first assess the contribution from emission directionality using FDTD simulations of the full and half-cavity configurations. The corresponding far-field intensity distributions are shown in Fig.~\ref{Pl}(d), with the  full and half-cavity results displayed on the left and right, respectively. In the half-cavity configuration, the emission extends over a wide angular range with an emission angle of approximately 50$^\circ$.  In contrast, the full-cavity configuration confines the emission to a much narrower angular range of approximately 5$^\circ$. The narrower angular distribution results from the directional selectivity of the resonant cavity mode. In the full cavity, the concave mirror provides transverse optical confinement, as discussed earlier, and supports a more directional far-field emission profile. The emission is therefore concentrated into a narrower angular cone, increasing the fraction of photons collected within the detection numerical aperture. 

% \subsection{Lifetime modification in cavity}
\textit{\textbf{Lifetime modification:}}  Having established improved emission directionality as one contribution to the enhanced PL intensity, we next examined whether the cavity also modifies the spontaneous-emission dynamics. For this purpose, TRPL  measurements were performed for both cavities, and corresponding decay traces are compared in Fig. \ref{Lifetime}(a). The lifetimes measured for the plano-concave cavities with lens diameters of 4, 5 and 6~$\mu$m are compared in Supplementary Fig.~\ref{Liftime-all}. Among these configurations, the 5~$\mu$m lens exhibits the largest reduction in emission lifetime relative to the half-cavity reference, consistent with the highest PL enhancement observed for this cavity. 

In the half-cavity configuration, the decay is well described by a bi-exponential function with two characteristic lifetimes, $\tau_0 = 270$ ps  and $\tau_1 = 961$ ps. The plausible origin of these two decay components, including their relation to different orientations of the $\mathrm{V_B^-}$ transition dipoles, is discussed in the following section. When the cavity resonance is tuned into spectral overlap with the $\mathrm{V_B^-}$ emission, both decay components become shorter. The fast component decreases to $\tau_0 = 45$ ps, while the slower component is reduced to $\tau_1 = 720$ ps. For the fast component, the lifetime is shortened by a factor of approximately 6. The corresponding Purcell factor can be estimated from the lifetime ratio:
\begin{equation}
F_{\mathrm{P}} =
\frac{\tau_{\mathrm{half}}}{\tau_{\mathrm{full}}}.
\end{equation}
Using the measured lifetimes, we obtain $F_{\mathrm{P}} \approx 6$. In contrast, the longer lifetime component shows only a weak reduction from 961 ps to 720 ps, corresponding to a Purcell factor of approximately 1.35. All decay curves were fitted using a convoluted model (see Supplementary Section~\ref{sec:IRF}) with the measured instrument response function (IRF), as detailed in Supplementary Fig.~\ref{IRF}.

The cavity-length dependence of the lifetime was further measured by tuning the resonance away from the defect emission, as shown in Fig.~\ref{Lifetime}(b). As the cavity mode is detuned, the local density of optical states changes according to the spectral overlap between the emitter and the cavity resonance. This behaviour can be described by the Lorentzian cavity density of states 
\cite{Wang2021,Munsch2009},
\begin{equation}
\rho_{\mathrm{cav}}(\omega)
\propto
\frac{(\kappa/2)^2}
{(\omega-\omega_c)^2+(\kappa/2)^2},
\end{equation}
where $\omega_c$ is the cavity resonance frequency and $\kappa$ is the full width at half maximum (FWHM) of the cavity resonance, related to the quality factor by $Q=\omega_c/\kappa$. The measured lifetime exhibits a resonance-dependent behaviour, with the strongest shortening observed near the cavity resonance and a gradual recovery as the cavity is detuned. This trend confirms that the lifetime modification is governed by the cavity-controlled photonic environment.

The FDTD calculations further predict the spectral dependence of the Purcell enhancement, as shown in Fig.~\ref{Lifetime}(c). The calculated Purcell factor for the in-plane dipole varies across the $\mathrm{V_B^-}$ emission band and reaches a maximum of approximately 7 near 840~nm. Away from resonance, the Purcell enhancement decreases as the spectral overlap with the cavity mode is reduced. The simulated maximum is in good agreement with the shortening of the fast component of the experimentally observed lifetime.

\subsection{Dipole orientation and polarisation study}
\begin{figure*}[t]
%\begin{strip}
\centering
    \includegraphics[width=1.0\textwidth]{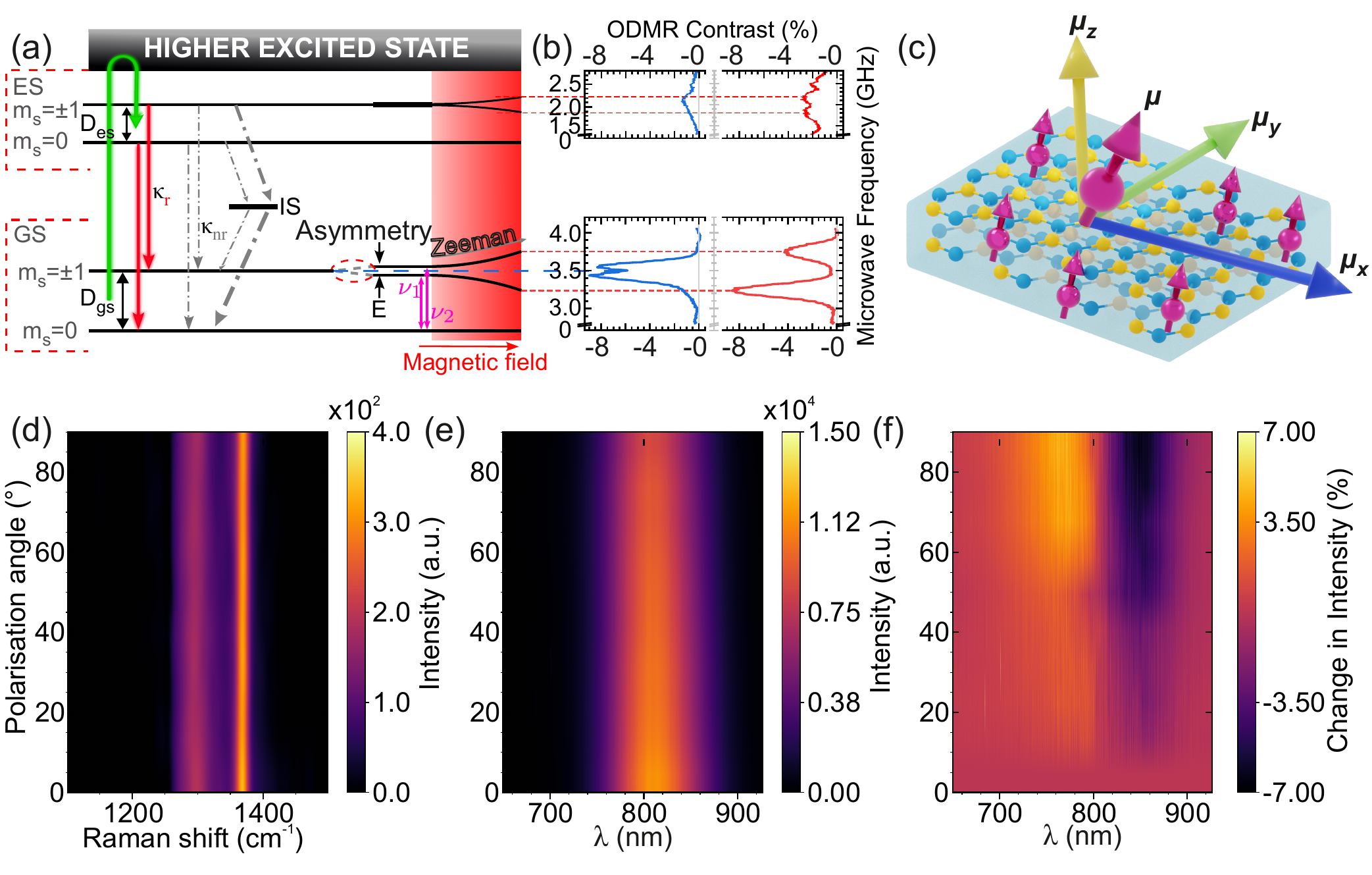}
    \caption{\textbf{Spin structure and polarisation response of the $\mathrm{V_B^-}$ defect emitter in hBN.}
    \textbf{(a)} Schematic representation of the relaxation pathways associated with the $\mathrm{V_B^-}$ centre, including optical excitation, radiative emission, and spin-dependent non-radiative relaxation channels. The level scheme also shows the ground-state and excited-state spin structure, including asymmetry-induced splitting and Zeeman splitting under an applied magnetic field.
    \textbf{(b)} Continuous-wave ODMR spectra of the ground and excited states. The observed resonance features confirm the spin-active nature of the boron-vacancy centres and show the expected splitting of the spin sublevels, including clear Zeeman shifts under an applied magnetic field.
    \textbf{(c)} Schematic illustration of the optical transition dipole orientation in the hBN crystal, showing the in-plane and out-of-plane dipole components.
    \textbf{(d)} Polarisation-resolved Raman spectra showing the $D_1$ and $E_{2g}$ modes at   1299.62~$\pm$~0.93~cm$^{-1}$ and 1370.20~$\pm$~0.16~cm$^{-1}$  with nearly constant intensity upon rotation of the incident polarisation.
    \textbf{(e)} Polarisation-resolved PL  showing an approximately 24\% reduction in emission intensity upon changing the incident polarisation from $0^\circ$ to $90^\circ$. 
    \textbf{(f)} Differential PL spectra plotted as a function of polarisation angle, revealing an asymmetric polarisation response of the defect emission. }
    \label{ODMR-Pol}
\end{figure*}

The observed cavity response motivates a closer examination of the $\mathrm{V_B^-}$ emitter itself, particularly the symmetry and orientation of its optical transitions. The $\mathrm{V_B^-}$ defect possesses $D_{3h}$ symmetry, although local strain and structural distortions can reduce its symmetry to $C_{2v}$. The asymmetry observed in ODMR measurements has provided experimental indications of symmetry lowering towards $C_{2v}$\cite{Mathur2022}. Such symmetry reduction can theoretically allow out-of-plane-polarised optical transitions \cite{Ivdy2020}. However, the orientation of the $\mathrm{V_B^-}$ emission dipole remains under debate in the literature \cite{Froch.2021, Sortino2024}. Here, we discuss the $\mathrm{V_B^-}$ dipole orientation by combining ODMR and polarisation-resolved PL measurements, which  helps to explain the cavity response discussed in the previous section.
The relaxation scheme in Fig.~\ref{ODMR-Pol}(a) shows the relevant optical and spin-dependent relaxation pathways. The 
system contains a triplet GS (with spin sublevels $m_\mathrm{s} = 0$ and $m_\mathrm{s} = \pm 1$, separated by a ZFS of \SI{3.5}{\GHz}. The splitting of the excited state (ES) is approximately \SI{2.09}{\GHz} \cite{Baber.2022}. In addition, a singlet intermediate state, also commonly called the metastable state, provides a non-radiative relaxation channel.

Under optical excitation, the $\mathrm{V_B^-}$ centres are promoted to higher-energy states and then rapidly relax to the ES through internal conversion. From the ES, non-radiative and phonon-assisted relaxation (denoted by $k_\mathrm{nr}$ and $k_\mathrm{r}$, respectively) to the GS occurs \cite{Reimers2020}, while the latter gives rise to broadband PL in the near-infrared spectral range. 
These optical transitions are predominantly spin-conserving, but their oscillator strength is weak because of the symmetry-forbidden nature of the transition, resulting in intrinsically low emission intensity. 
Relaxation through the ISC is  spin-dependent, where $m_\mathrm{s}=\pm1$ has roughly half the lifetime of the $m_\mathrm{s}=0$ transition \cite{Baber.2022, CluaProvost.2024}. The subsequent reverse ISC has a short lifetime and favours relaxation into the ground-state $m_\mathrm{s}=0$ state, leading to spin polarisation under continuous excitation \cite{Konrad.2026,Gottscholl2020}.
The ZFS parameters $D$ and $E$ can be determined by ODMR (see blue curves in Fig.~\ref{ODMR-Pol}(b)) for GS and ES transitions).
The GS $m_s = \pm 1$ sublevels show a ZFS with $D=\SI[round-mode=uncertainty,round-precision=1, separate-uncertainty=true]{3491 \pm 0.8598}{\mega\hertz}$ and $E =\SI[round-mode=uncertainty,round-precision=1, separate-uncertainty=true]{74\pm 1}{\mega\hertz}$ (for details see Supplementary Fig.~\ref{fig:SI:CPW}, \ref{fig:SI:Flake}, \ref{fig:SI:ZFS-GS}). 
Under an external magnetic field of {\SI[round-mode=uncertainty,round-precision=1, separate-uncertainty=true]{8.69\pm 0.03406}{\milli\tesla}}, the Zeeman interaction increases the splitting (see red curve in Fig.~\ref{ODMR-Pol}(b)). 
The ES ODMR spectrum, shown in the upper panel of Fig.~\ref{ODMR-Pol}(b), does not show any resolved transverse zero-field splitting, which may be because it is within the ODMR linewidth. When an external magnetic field is applied, the excited-state resonance splits into two branches, as expected for Zeeman splitting of the spin-triplet sublevels.
In this picture, the defect optical transition dipole in the GS contains some asymmetry, and we hypothesise both out-of-plane and in-plane components, as schematically illustrated in Fig.~\ref{ODMR-Pol}(c).
 
Polarisation-resolved Raman and PL spectroscopy were performed to probe the character of the  optical dipole of the $\mathrm{V_B^-}$ emission, as shown in Fig.~\ref{ODMR-Pol}(d-f). 
The polarisation direction of the incident laser light is varied using a half-wave plate ($\lambda$/2) placed in the excitation path. During the measurement, an analyser is placed in the detection path to select the horizontally polarised component of the scattered light.

The intensities of the $D_1$ and $E_{2g}$ Raman modes remain nearly unchanged as a function of polarisation angle, as shown in Fig.~\ref{ODMR-Pol}(d). The expected isotropic behaviour of the in-plane Raman modes \cite{Patra2025} indicates that setup-related artefacts do not dominate the measurement. The corresponding polarisation-resolved PL spectra are shown in Fig.~\ref{ODMR-Pol}(e). A reduction of approximately 24\% in the PL intensity is observed upon rotating the incident light polarisation, indicating a polarisation-dependent optical response of the defect emission. This behaviour suggests the presence of an anisotropy in the dipole component and provides confirmation of the tilted dipole. This behaviour suggests anisotropic dipole orientation, consistent with a tilted dipole rather than one aligned perpendicular to the hBN plane.
To visualise the spectral changes more clearly, the differential PL was calculated from spectra scaled to their highest intensities at each excitation-polarisation angle, as shown in Fig.~\ref{ODMR-Pol}(f):
\begin{equation}
\Delta I(\lambda,\theta)
=
N\!\left[I(\lambda,\theta)\right]
-
N\!\left[I(\lambda,\theta_{\mathrm{ref}})\right],
\end{equation}
where N means normalised intensity from 0 to 1, $\theta$ is the polarisation angle and $\theta_{\mathrm{ref}} = 0^\circ$ is the reference angle. Since the differential PL is calculated after normalisation, positive and negative spectral features directly represent changes in the relative emission intensity. Positive values indicate enhanced emission at a given wavelength compared with the reference polarisation, while negative values indicate reduced emission. The appearance of these polarisation-dependent positive and negative features across the $\mathrm{V_B^-}$ emission band supports an asymmetric dipole response of the defect centres. 

\section{Discussion}
We present the first study of $\mathrm{V_B^-}$ defects in hBN integrated with a tunable open microcavity, together with an experimental investigation of their optical dipole orientation. The tunable open-microcavity platform addresses the limitations of fixed cavity structures by allowing external control over the cavity length and resonance condition while avoiding direct nanofabrication of the active hBN flake. 
We investigate the light-matter interaction between the broadband $\mathrm{V_B^-}$ emission and the plano-concave cavity, where the observed response reflects the combined contributions of collection efficiency, emission directionality and the local density of optical states. 
The reduction in the TRPL lifetime is consistent with a modification of the local density of optical states arising from three-dimensional photonic confinement. However, the distinct changes in the biexponential decay components raise further questions about the interaction between the emitter and the confined cavity field, motivating an investigation of the optical dipole orientation. 
The ODMR response of $\mathrm{V_B^-}$ defects in hBN for both GS and ES can be described by the spin Hamiltonian
\begin{equation}
H = hD\left(S_z^2 - \frac{S(S+1)}{3}\right)
+ hE\left(S_x^2 - S_y^2\right)
+ g\mu_{\mathrm{B}}\mathbf{B}\cdot\mathbf{S}
\label{eq:spin_hamiltonian}
\end{equation}
where $D$ and $E$ represent the ZFS parameters, h is Planck’s constant, and $\mathbf{S}$ is the total spin operator with its components $S_i$ in direction $i$, $S$ the (dimensionless) total spin number. The last term describes an electron Zeeman interaction with  the electron g-factor $g$,  the Bohr magneton $\mu_{\mathrm{B}}$ and  an external magnetic field $\mathbf{B}$. 
The axial parameter $D$ is compatible with the ideal $D_{3h}$ symmetry, whereas a finite $E$ lift the degeneracy of the $m_s=\pm1$ sublevels, indicating a change from the ideal threefold symmetry, consistent with symmetry lowering towards $C_{2v}$ \cite{Mathur2022}. 
Deviations can arise from local symmetry breaking, such as strain in the hBN lattice, inequivalent defect orientations, or crystal-field asymmetry. This indicates that the defect's local environment is not perfectly axial with respect to the \textcolor{blue}{\textit{c}}-axes. 
%moved from experimental part
Together with the polarisation-dependent optical response, this motivates a dipole geometry in which the transition dipole is not purely out of plane but is tilted with respect to the hBN crystal axis. 
In this picture, the defect dipole contains both out-of-plane and in-plane components, as schematically illustrated in Fig.~\ref{ODMR-Pol}(c).
The observed ground-state splitting is consistent with such anisotropy, whereas the absence of a resolved excited-state splitting suggests a smaller or spectrally unresolved $E$ parameter. We can only estimate $E$ to be close to 0 for the ES (for further details, see Supplementary Section \ref{sec:ODMR}).
For the GS, we find an asymmetry of {\SI[round-mode=uncertainty,round-precision=1]{2.068 \pm 0.0141}{\percent}}. %how does is affect the PL
These observations suggest state-dependent spin anisotropy associated with local symmetry-breaking perturbations, which may also influence the optical selection rules  \cite{Ivdy2020}.

In further investigation, we observe a spectral separation of two components depending on the polarisation angle, which indicates a contribution from an out-of-plane and an in-plane optical dipole component. Here, the out-of-plane component is the biggest contributor to PL at lower energies.
The increase of PL at \SI{840}{\nm} can be attributed to an in-plane component in this wavelength range.

Returning to the lifetime measurements, this interpretation is consistent with the different responses of the two PL decay components. The in-plane dipole component can couple efficiently to the confined cavity field, consistent with the pronounced reduction in $\tau_0$, whereas $\tau_1$ exhibits only a modest change. The smaller modification of $\tau_1$ may be associated with differences in emitter-field overlap, imperfections in the fabricated lens geometry or changes in the relaxation dynamics under continuous optical excitation. However, the individual contributions of these mechanisms cannot be distinguished from the present measurements.

In summary, we demonstrate the integration of $\mathrm{V_B^-}$ spin defects in hBN with a tunable open microcavity, enabling controlled spectral coupling between broadband defect emission and confined optical resonances. Beyond modifying the emission characteristics, our combined TRPL, ODMR and polarisation-dependent measurements provide insight into the interplay between the local defect environment, optical transition dipole orientation and cavity-mediated light--matter interaction. These findings establish tunable open microcavities as a versatile platform for investigating and controlling the optical response of spin defects in van der Waals materials. %More importantly, it provides an approach to enhance the QY of $\mathrm{V_B^-}$ centres.

\section{Acknowledgement--} 
A.K.S., A.P., P.K., S.H., and V.D. acknowledge funding from the lighthouse project IQ-Sense of the Bavarian State Ministry of Science and the Arts as part of the Bavarian Quantum Initiative Munich Quantum Valley (15 02 TG 86). We acknowledge ﬁnancial support from the Würzburg-Dresden Cluster of Excellence on Complexity and Topology in Quantum Matter ctd.qmat (EXC 2147, DFG project ID 390858490). V.D. acknowledges the support by the European Research Council (ERC) (Grant agreement No. 101055454). We acknowledge support from the Ion Beam Centre (IBC) at Helmholtz-Zentrum Dresden-Rossendorf (HZDR) for ion implantation. A.K.S. and A.P. thank Christian Schneider's group for discussing the open-cavity setup. 
%\clearpage
% Print references for the main manuscript
\printbibliography[
    heading=bibintoc,
    title={References}]

\end{refsection}

% Supplementary information
\ifwithsupplementary
    \clearpage
    \newgeometry{margin=2.2cm}
    \onecolumn

% When this file is compiled by itself, switch from the two-column parent
% class to a full-width, single-column SI layout. In the combined manuscript,
% main.tex performs this switch before loading the subfile.
\ifSubfilesClassLoaded{%
  \newgeometry{margin=2.2cm}%
  \onecolumn
}{}

% This refsection keeps supplementary citations and references separate from
% the main-paper bibliography. Put all SI-only BibTeX entries in SI_refs.bib.
\begin{refsection}[SI_refs.bib]
% Independent numbering for the supplementary information.
\setcounter{section}{0}
\setcounter{subsection}{0}
\setcounter{figure}{0}
\setcounter{table}{0}
\setcounter{equation}{0}
\setcounter{secnumdepth}{2}
\renewcommand{\thesection}{S\arabic{section}}
\renewcommand{\thesubsection}{S\arabic{section}.\arabic{subsection}}
\renewcommand{\thefigure}{S\arabic{figure}}
\renewcommand{\thetable}{S\arabic{table}}
\renewcommand{\theequation}{S\arabic{equation}}

% The main manuscript suppresses section numbers; restore them locally for SI.
\titleformat{\section}
  {\normalfont\bfseries\large\raggedright}
  {\thesection}
  {0.6em}
  {}
\titleformat{\subsection}
  {\normalfont\bfseries\normalsize\raggedright}
  {\thesubsection}
  {0.6em}
  {}

\begin{center}
  {\LARGE\bfseries Supplementary Information\par}
  \vspace{0.7em}
  {\large Cavity Control of Emission Dynamics from Spin Defects in Hexagonal Boron Nitride\par}
  \vspace{0.7em}
  {\normalsize Anuj Kumar Singh et al.\par}
\end{center}

\vspace{1em}

 \section{Sample fabrication}
\subsection{Substrate identification for the bottom and top DBR mirrors}
\label{sec:substrate}
The intrinsically low quantum yield of $\mathrm{V_B^-}$ defects makes their emission particularly susceptible to background photoluminescence. Careful selection of the cavity substrates is therefore crucial to suppress unwanted emission from the mirrors within the spectral range of $\mathrm{V_B^-}$, ensuring reliable detection of the defect photoluminescence. We therefore first screened the emission of different substrate materials before fabricating the bottom and top DBR mirrors. The PL spectra of possible bottom-mirror substrates are shown in Supplementary Fig.~\ref{substrate}(a-c). GaAs shows a pronounced PL contribution, while glass exhibits autofluorescence that overlaps with the broadband spectral window of the $\mathrm{V_B^-}$ defects. In contrast, Si shows negligible autofluorescence in this spectral region. Based on this comparison, Si was selected as the substrate for the bottom mirror.
After substrate selection, we etched the mesa structure into the Si substrate, followed by sputtering of DBR mirros (see section \ref{sec:cavityalign}). This geometry provides a low-background bottom mirror while maintaining a well-defined platform for transferring the irradiated hBN flake.

For the top mirror, the substrate needed to be optically transparent. Because glass showed measurable autofluorescence in the relevant emission window, we chose SiC as the transparent top-mirror substrate. The emission comparison between glass and SiC is shown in Supplementary Fig.~\ref{substrate}(d), where SiC provides a lower background in the spectral range of interest. A concave lens structure was then milled into the SiC substrate using focused ion beam fabrication, followed by deposition of four SiO$_2$/TiO$_2$ DBR  mirrors (see section \ref{sec:cavityalign}). This combination of a low-background Si bottom mirror and a transparent SiC curved top mirror provides a suitable cavity platform for coupling to the broadband $\mathrm{V_B^-}$ emission.  

\begin{figure}[h]
    \centering
    \includegraphics[width=\textwidth]{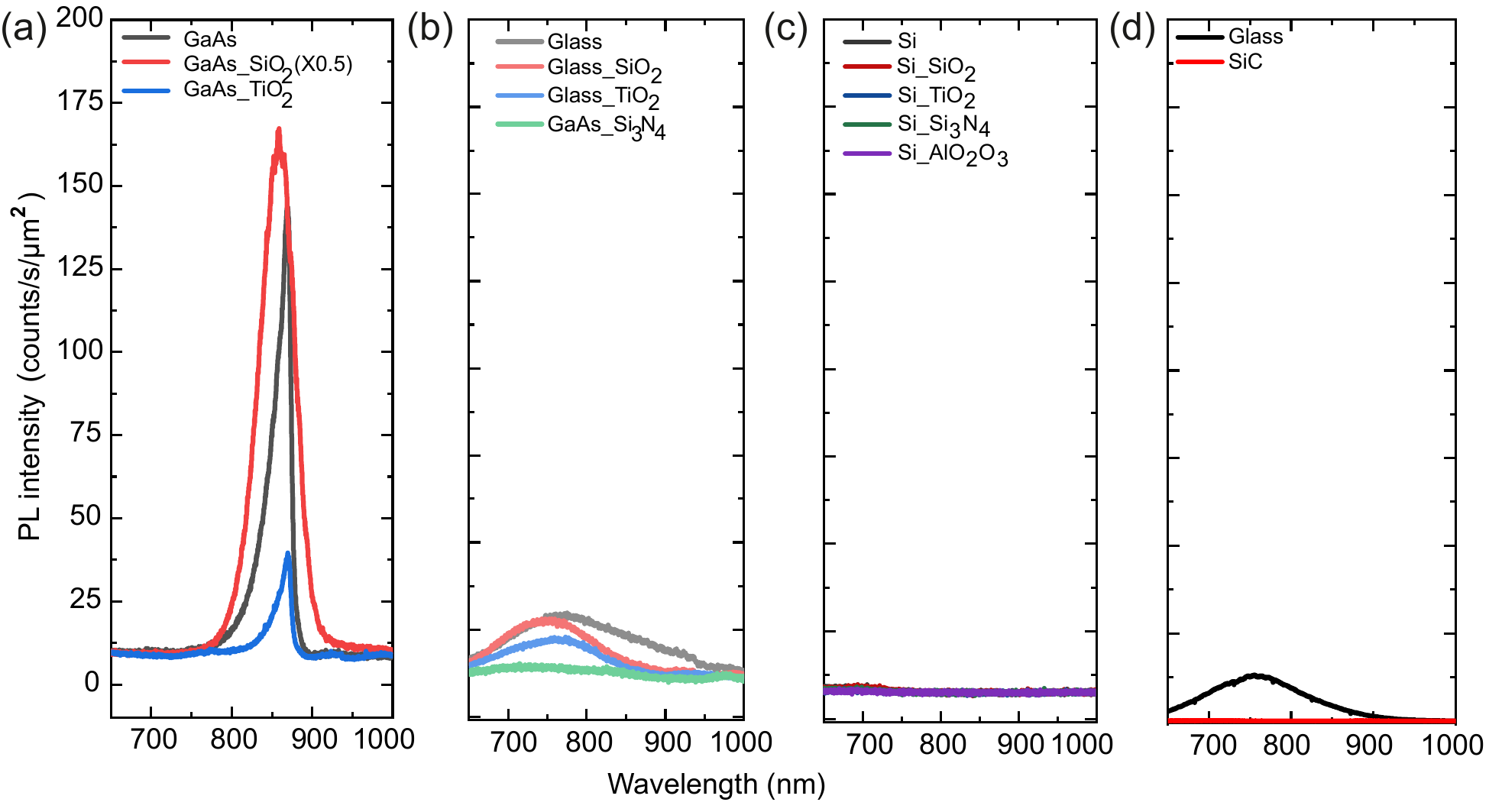}
    \caption{\textbf{Photoluminescence background comparison of substrates for open-cavity mirror fabrication.}
\textbf{(a)} Photoluminescence (PL) spectrum of the bare GaAs substrate and the SiO$_2$/TiO$_2$-coated GaAs substrate.
\textbf{(b)} PL spectrum of the bare glass substrate and the corresponding SiO$_2$/TiO$_2$-coated glass substrate.
\textbf{(c)} PL spectrum of the bare Si substrate and the SiO$_2$/TiO$_2$-coated Si substrate.
\textbf{(d)} PL comparison between glass and SiC substrates, highlighting their background emission of glass.} 

    \label{substrate}
\end{figure}
 \begin{figure}[h]
    \centering
    \includegraphics[width=\textwidth]{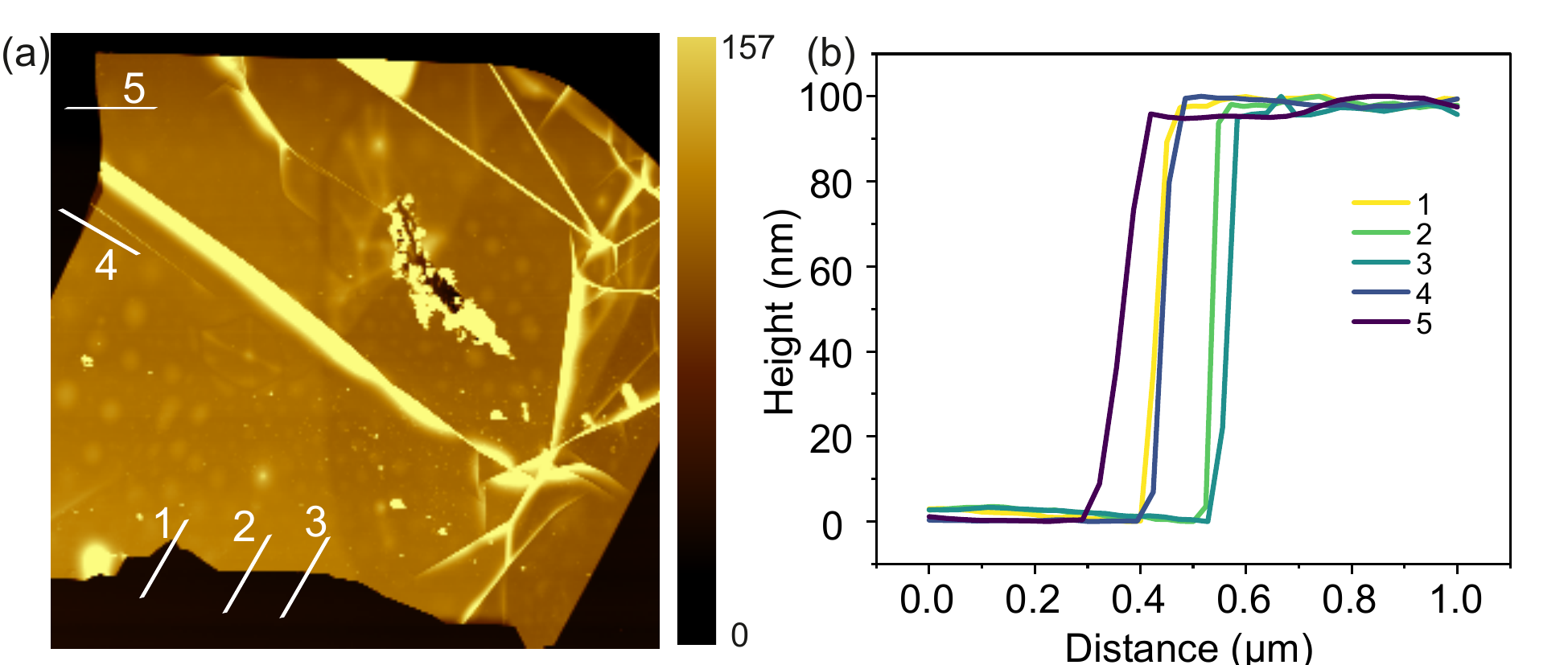} 
\caption{\textbf{AFM thickness profile of the hBN flake.}
\textbf{(a)} Atomic force microscopy (AFM) image of the selected hBN flake used for cavity integration.
\textbf{(b)} Height profile extracted along the marked line cut in panel (a), showing a flake thickness of approximately 100~nm.}
    \label{AFM}
\end{figure}
 
\subsection{Raman and PL map} 
\label{sec:RPL}
\begin{figure*}[h]
    \centering
    \includegraphics[width=\textwidth]{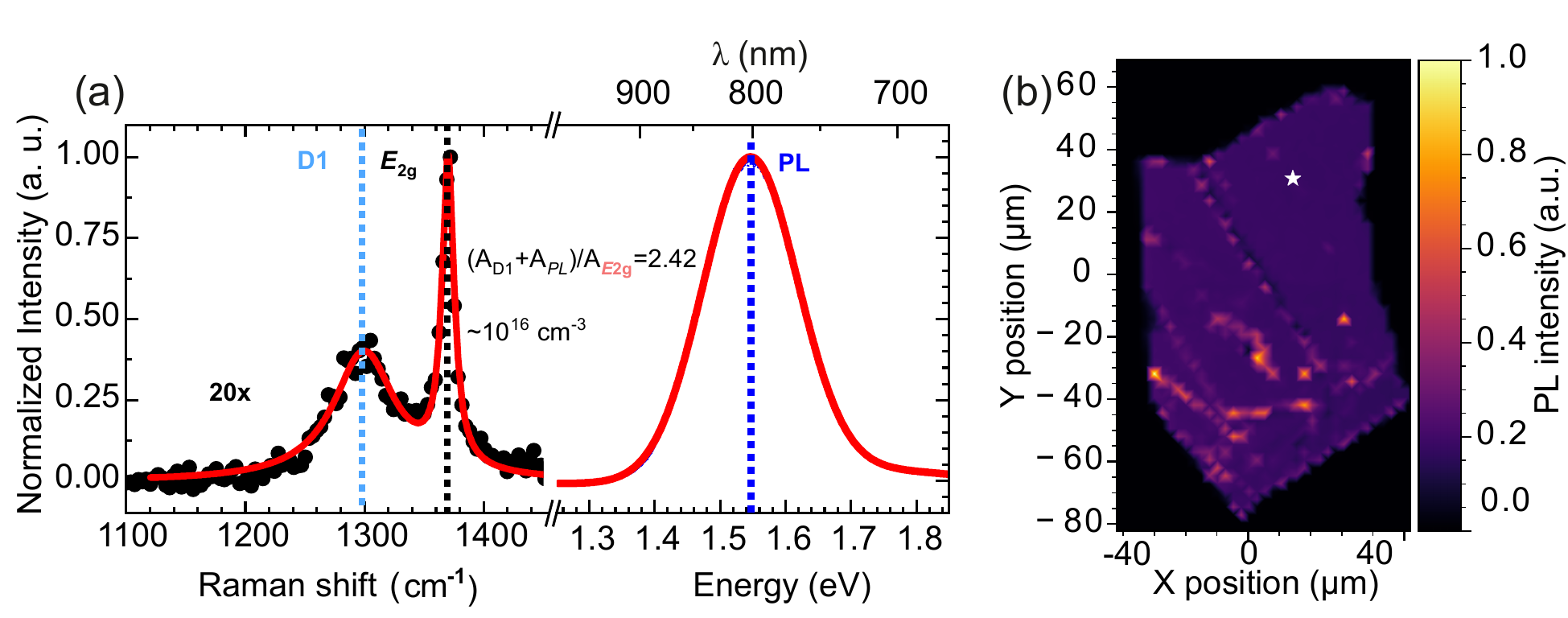} 
\caption{\textbf{Pre-characterisation of the flake.}
\textbf{(a)} Raman spectrum recorded from the hBN flake, showing defect-activated Raman features labelled as  $D_1$, together with the characteristic $E_{2g}$ in-plane optical phonon mode of hBN.
\textbf{(b)}Photoluminescence intensity map of the hBN flake, obtained by integrating the emission around 800~nm within a spectral window of $\pm 150$~nm. The spatially resolved map highlights the distribution of defect-related emission across the flake.}
    \label{Raman-Pl}
\end{figure*}
\begin{figure*}[h]
    \centering
    \includegraphics[width=\textwidth]{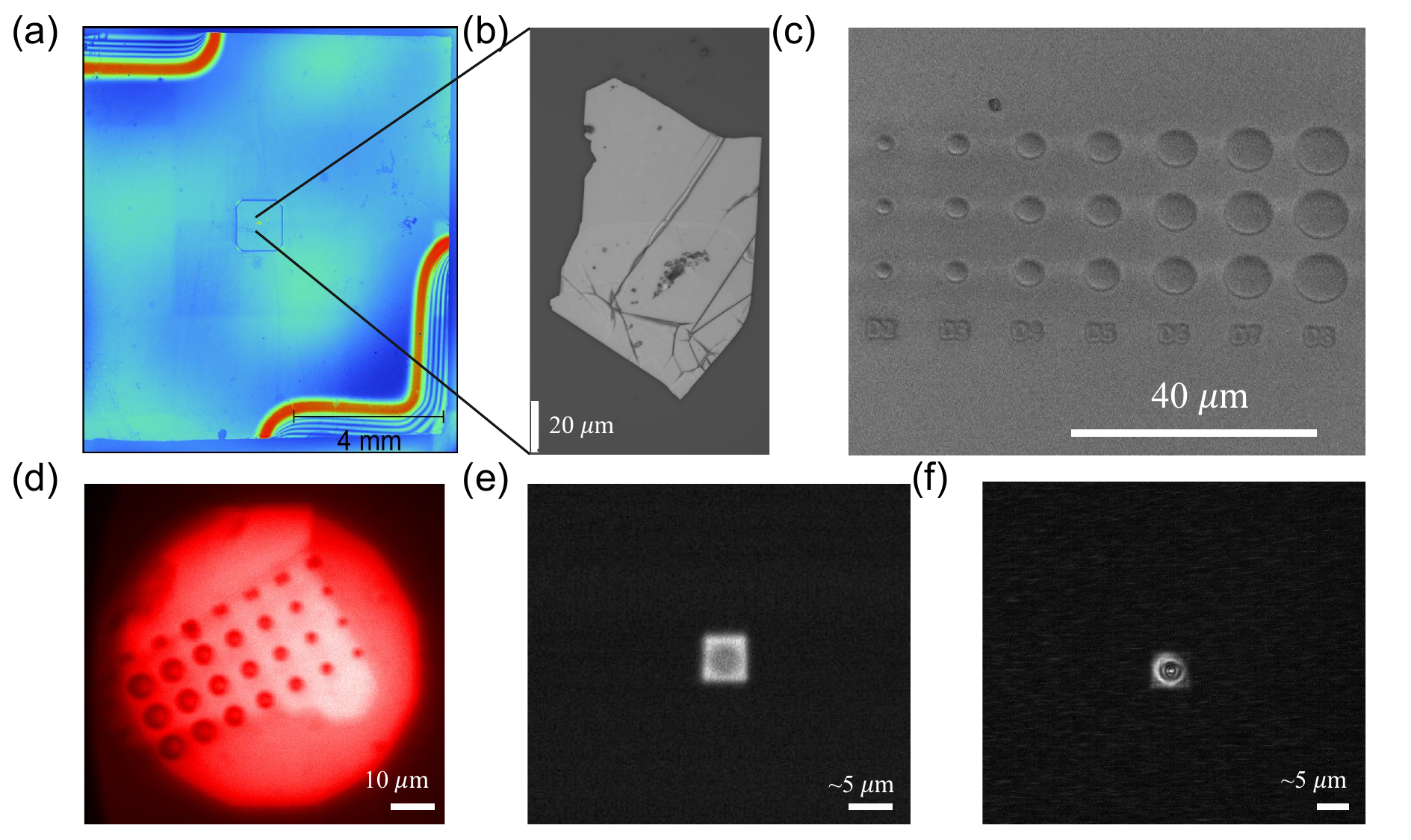}
      \caption{\textbf{Cavity design and alignment of the tunable open microcavity.}
\textbf{(a)} Laser scanning microscopy (LSM) image of the bottom mirror containing MESA structure (height  60~$\mu$m and dimension of 1 mm x 1 mm at the centre, onto which the pre-irradiated hBN flake was transferred.
\textbf{(b)} Optical microscopy image of the selected hBN flake used for cavity integration.
\textbf{(c)} Scanning electron microscopy (SEM) image of the  lens structure fabricated on the silicon carbide substrate by focused ion beam (FIB) milling.
\textbf{(d)} False-color optical image showing the top and bottom mirrors after coarse alignment and angle correction.
\textbf{(e)} Spatial filtering configuration used next to the $k$-space imaging lens to suppress scattered light from the cavity .
\textbf{(f)} Laser-spot alignment image showing the excitation spot positioned at the center of the lens on the top mirror.}
    \label{cavity_align}
\end{figure*}

\begin{figure*}[h]

    \centering
    \includegraphics[width=\textwidth]{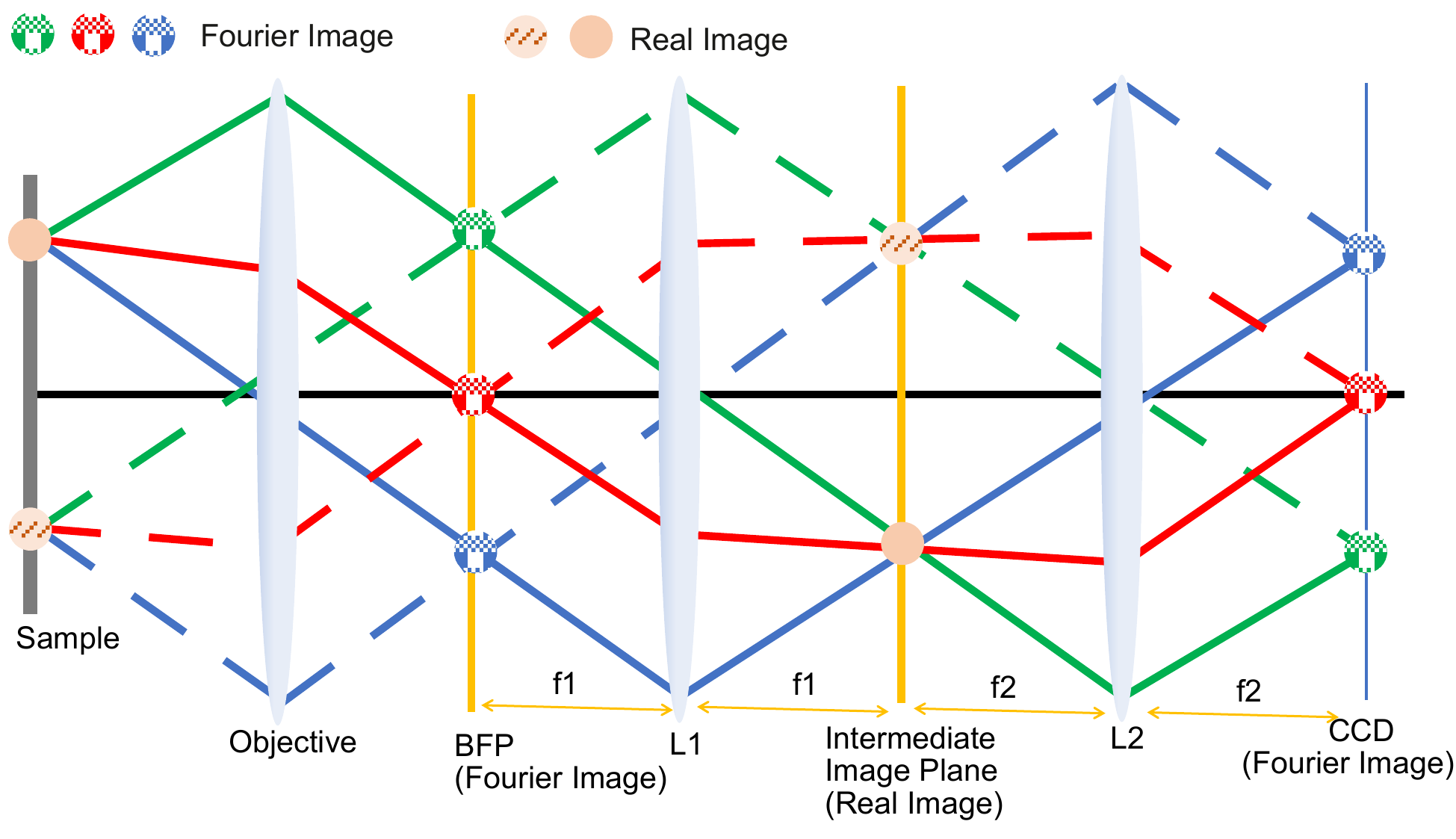}
      \caption{\textbf{Concept of the angle-resolved measurement using the Fourier lens}}
    \label{arm}
\end{figure*}

\begin{figure*}[h]
    \centering
    \includegraphics[width=\textwidth]{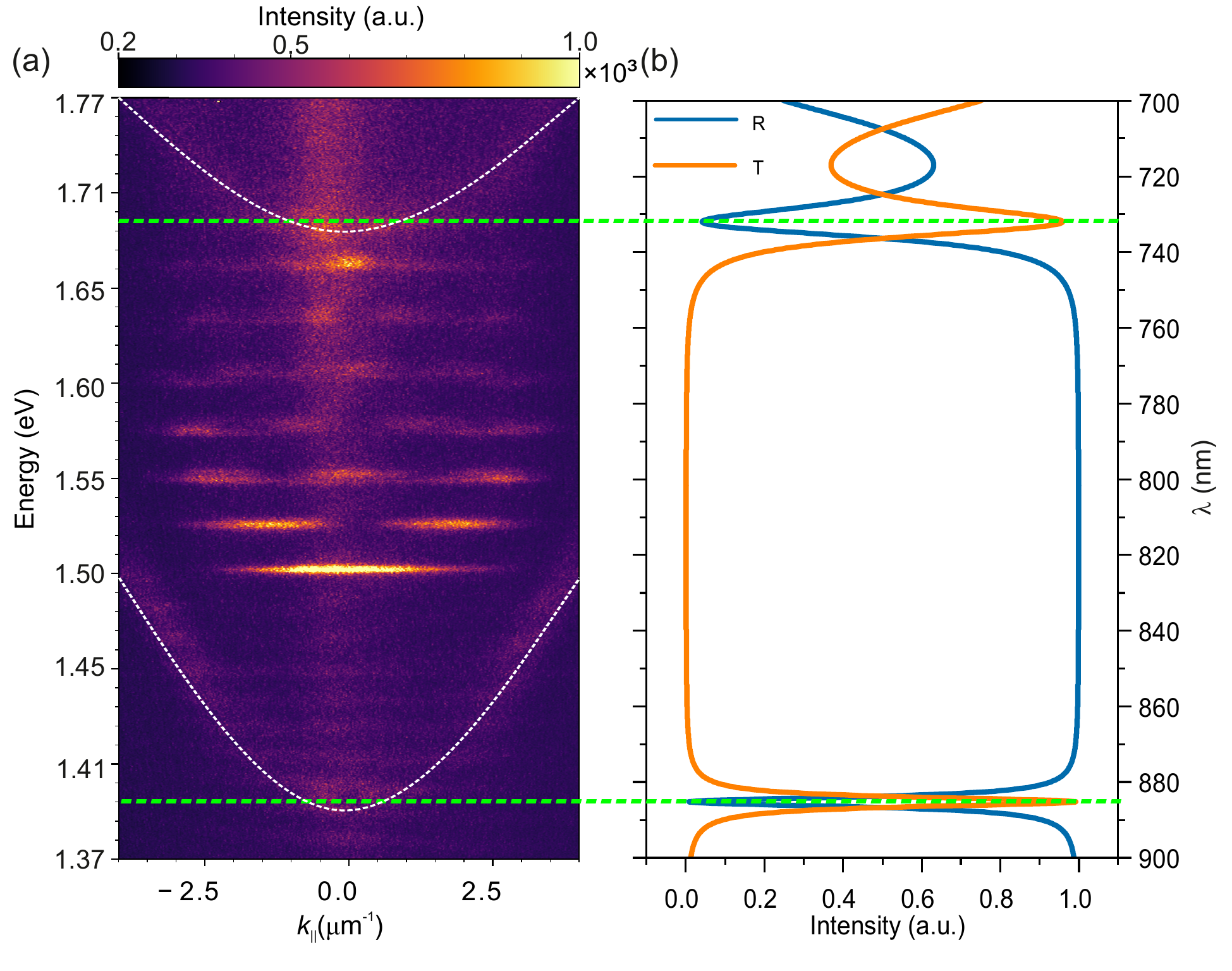}
    \caption{\textbf{Estimation of the actual air gap at the resonance position.}
\textbf{(a)} Experimental $k$-space image acquired at the resonance condition, showing two distinct cavity-mode families centered around 733~nm and 884~nm. 
\textbf{(b)} Transfer-matrix method (TMM) calculation performed by varying the air-gap thickness between the cavity mirrors. By matching the calculated cavity resonances with the experimentally observed mode families at 733~nm and 884~nm, the actual air gap at the resonance position is estimated to be approximately 720~nm.}
    \label{air-gap}
\end{figure*}

\begin{figure*}[h]
    \centering
    \includegraphics[width=\textwidth]{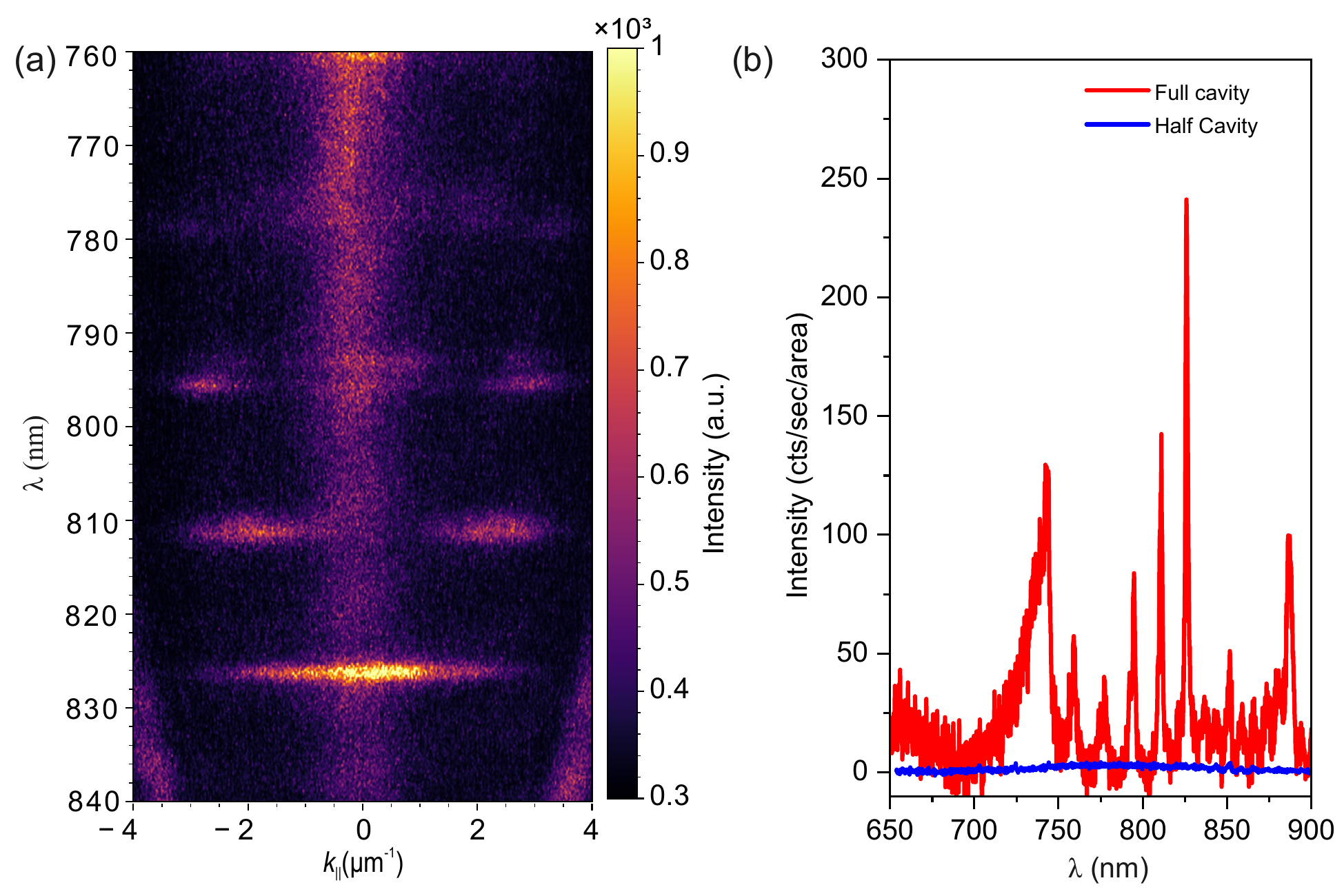}
  \caption{\textbf{Photoluminescence enhancement using a 4~$\mu$m-diameter  lens.} 
\textbf{(a)} Experimental $k$-space photoluminescence image showing coupling of the $\mathrm{V_B^-}$ defect emission to the cavity modes formed by the 4~$\mu$m-diameter  lens.
\textbf{(b)} Photoluminescence spectrum measured under the cavity-coupled condition compared with the corresponding half-cavity reference. An emission enhancement of approximately 77 times is observed for the 4~$\mu$m lens.}
    \label{D4}
\end{figure*}

\begin{figure*}[h]
    \centering
    \includegraphics[width=\textwidth]{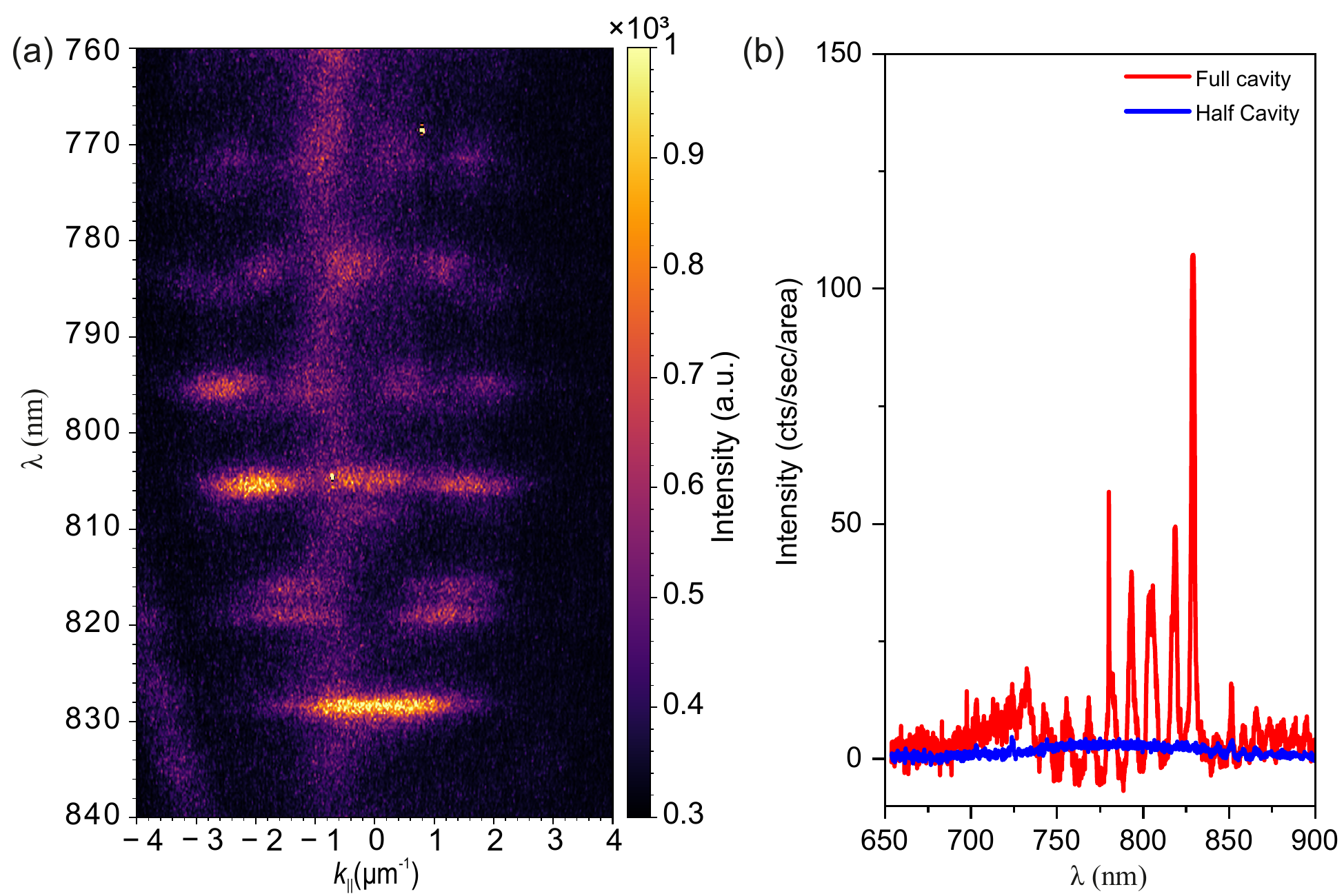}
 \caption{\textbf{Photoluminescence enhancement using a 6~$\mu$m-diameter  lens.}
\textbf{(a)} Experimental $k$-space photoluminescence image showing coupling of the $\mathrm{V_B^-}$ defect emission to the cavity modes formed by the 6~$\mu$m-diameter  lens.
\textbf{(b)} Photoluminescence spectrum measured under the cavity-coupled condition compared with the corresponding half-cavity reference. An emission enhancement of approximately 34 times is observed for the 6~$\mu$m lens.}
    \label{D6}
\end{figure*}

\begin{figure*}[h]
    \centering
    \includegraphics[width=\textwidth]{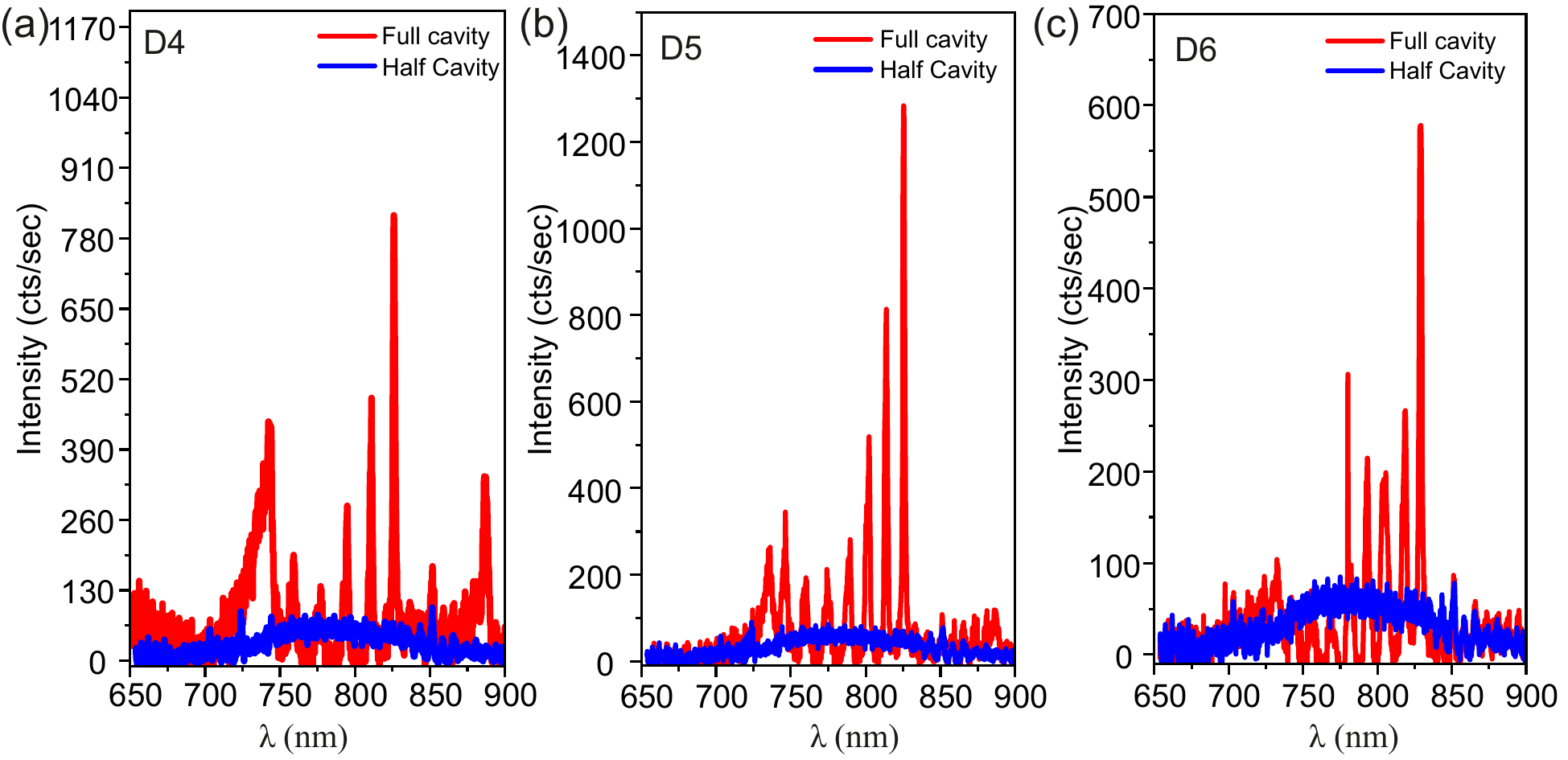}
\caption{\textbf{Raw photoluminescence data before normalisation.} 
\textbf{(a)} Raw PL spectra measured using a 4~$\mu$m-diameter  lens, showing an unnormalized enhancement factor of approximately 13.
\textbf{(b)} Raw PL spectra measured using a 5~$\mu$m-diameter lens, showing an unnormalized enhancement factor of approximately 20.
\textbf{(c)} Raw PL spectra measured using a 6~$\mu$m-diameter e lens, showing an unnormalized enhancement factor of approximately 9.}
    \label{PL raw}
\end{figure*}

\begin{figure*}[h]
    \centering
    \includegraphics[width=\textwidth]{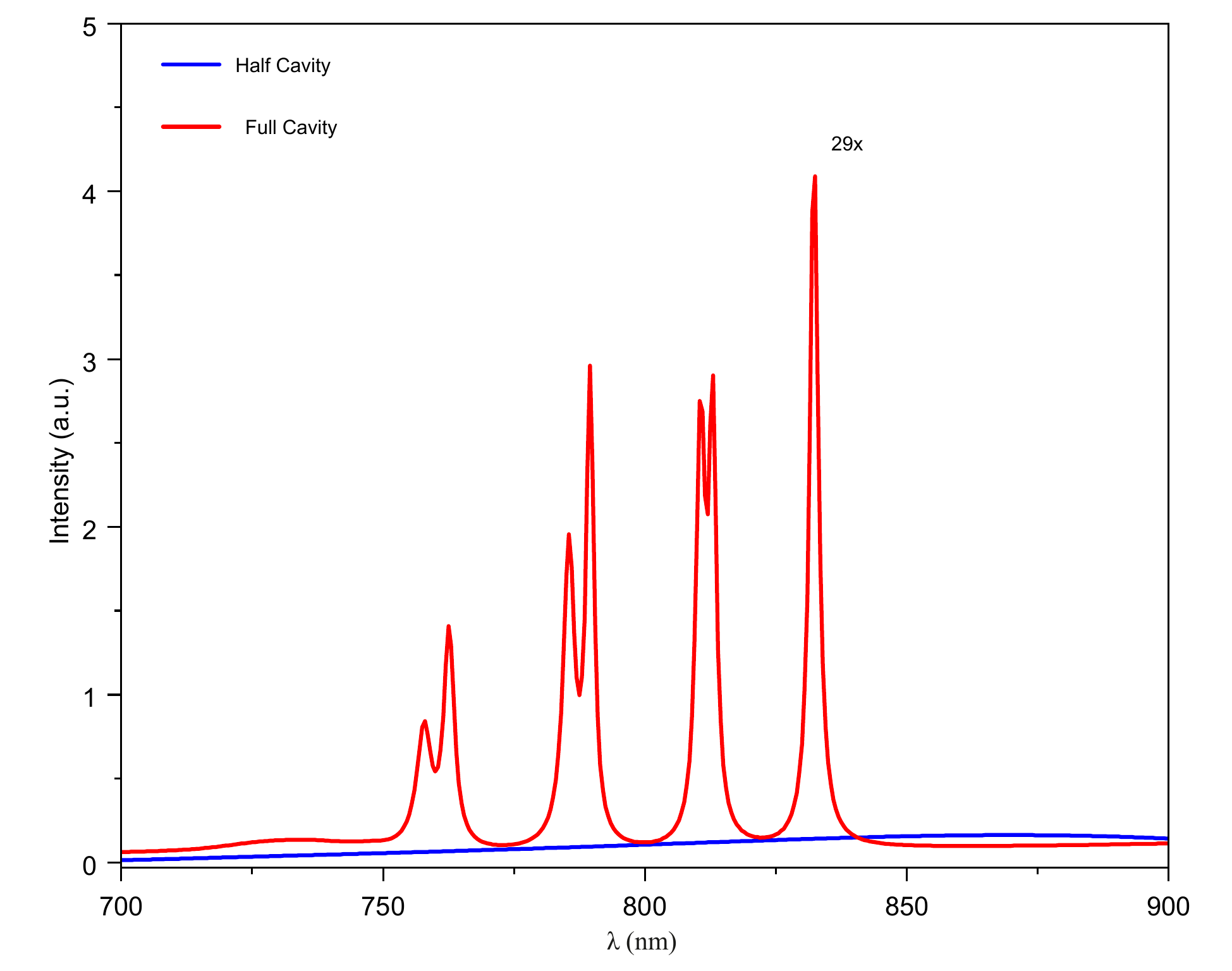}
\caption{\textbf{Lumerical Simulation.}
Transmission intensity in the half and full cavity gives the 29 times enhancement in the full  cavity compared to the half cavity }
    \label{T-FDTD}
\end{figure*}

\begin{figure*}[h]
    \centering
   \includegraphics[width=\textwidth]{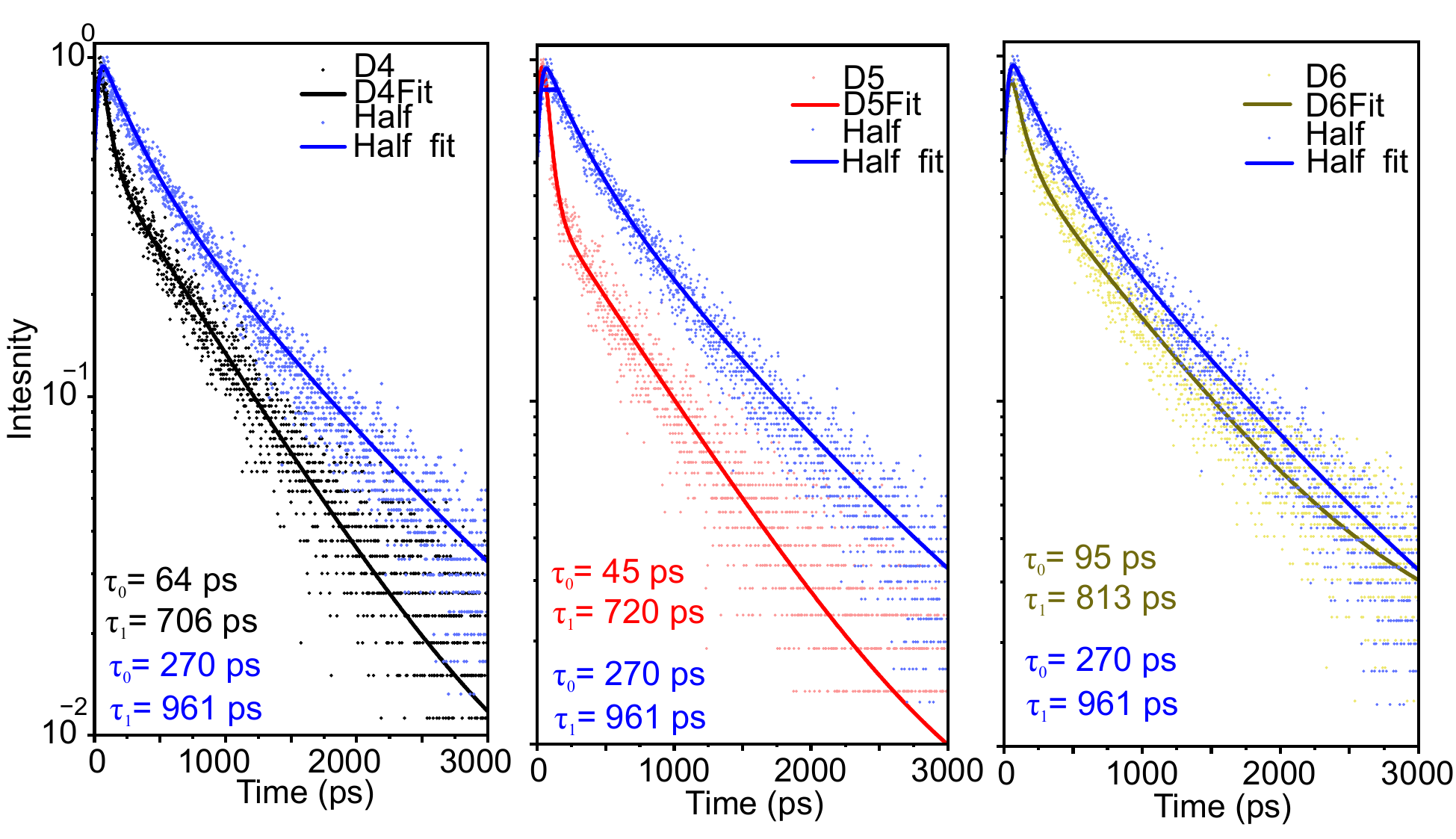}
\caption{\textbf{Lifetime measurements for cavities with different concave-lens diameters.}
\textbf{(a)} Time-resolved photoluminescence decay measured from the cavity-coupled defect emission using a 4~$\mu$m-diameter concave lens, 
\textbf{(b)} Lifetime decay trace and fit obtained for the cavity formed with a 5~$\mu$m-diameter concave lens.
\textbf{(c)} Lifetime decay trace and fit obtained for the cavity formed with a 6~$\mu$m-diameter concave lens.}
    \label{Liftime-all}
\end{figure*}

\begin{figure*}[h]
    \centering
    \includegraphics[width=\textwidth]{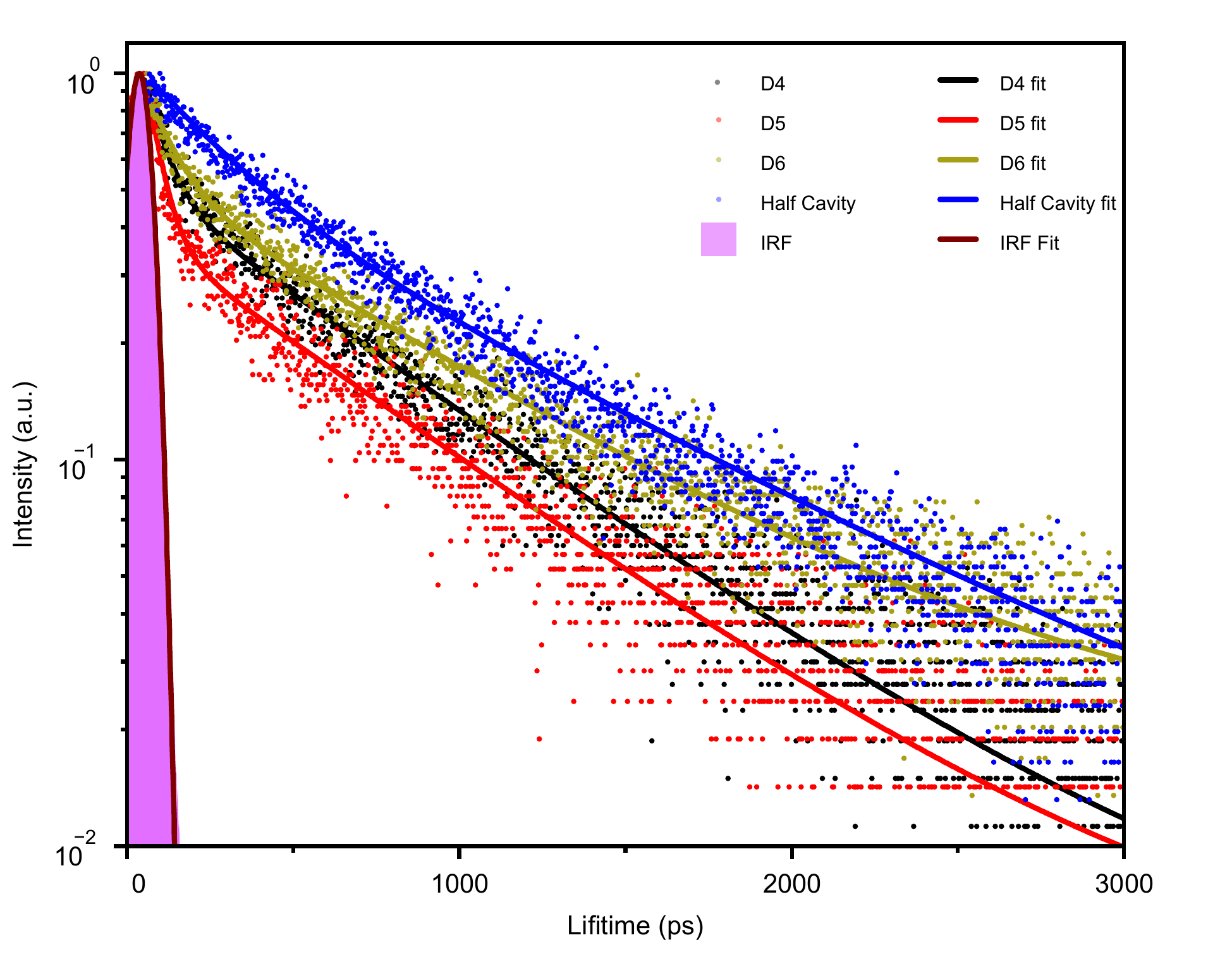}
\caption{\textbf{Instrument response function.}
IRF of the TCSPC system fitted with a Gaussian function. The fit yields a standard deviation of $\sigma = 34.9$~ps, corresponding to an IRF full width at half maximum (FWHM) of about $82.2$~ps, which was used to deconvolve the lifetime data.}
    \label{IRF}
\end{figure*}
\begin{figure}
    \centering
    \includegraphics[width=0.5\linewidth]{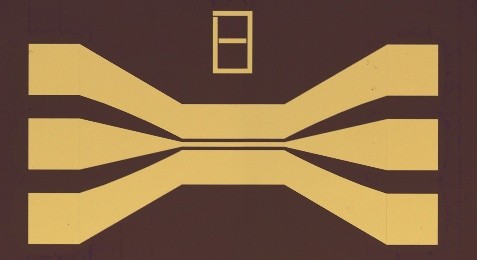}
    \caption{\textbf{Coplanar waveguide for ODMR.}}
    \label{fig:SI:CPW}
\end{figure}
\begin{figure}
    \centering
    \includegraphics[width=0.5\linewidth]{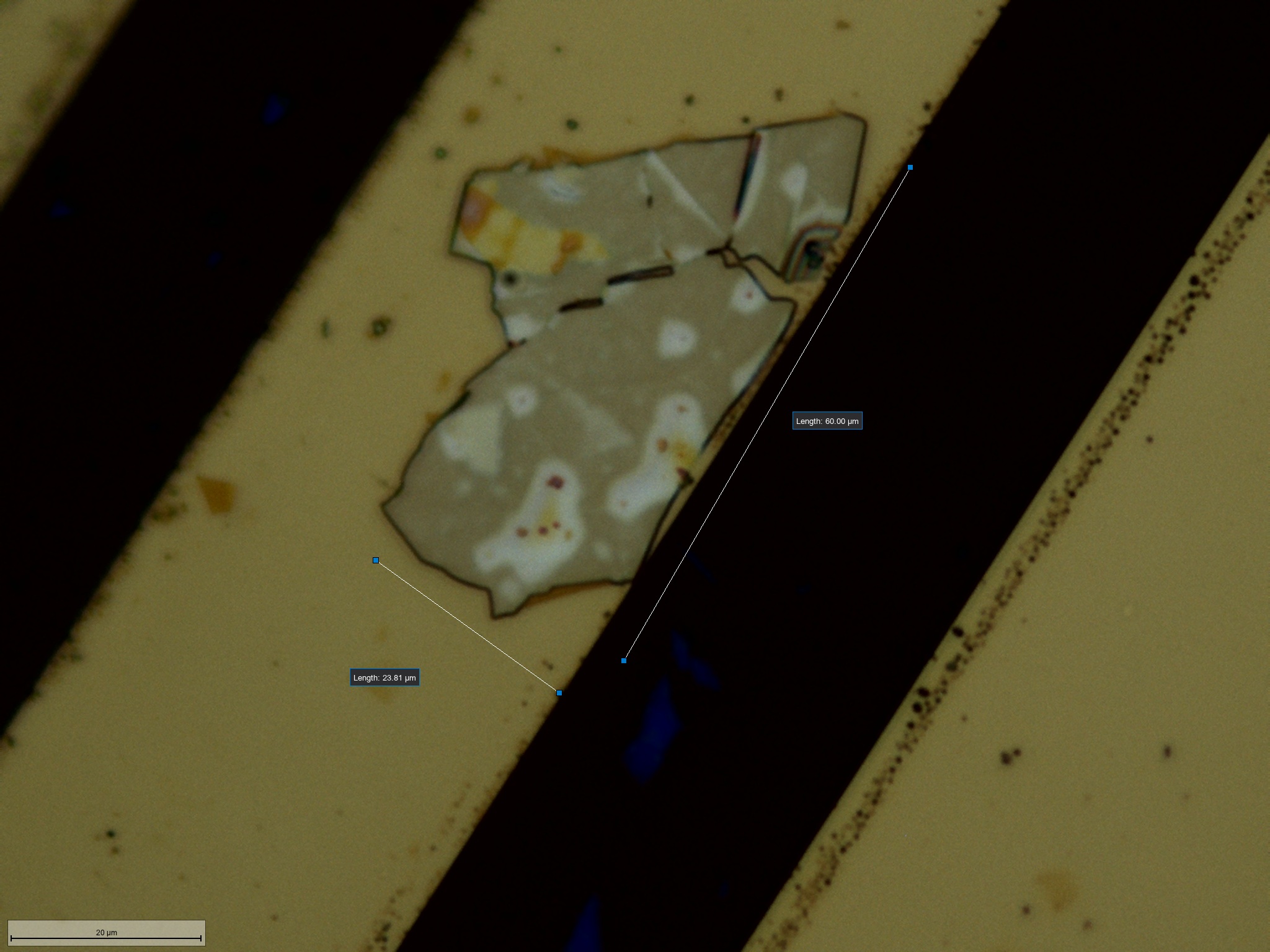}
    \caption{\textbf{Irradiated hBN flake on CPW.}}
    \label{fig:SI:Flake}
\end{figure}

\begin{figure*}[h]
    \centering
    \includegraphics[width=\textwidth]{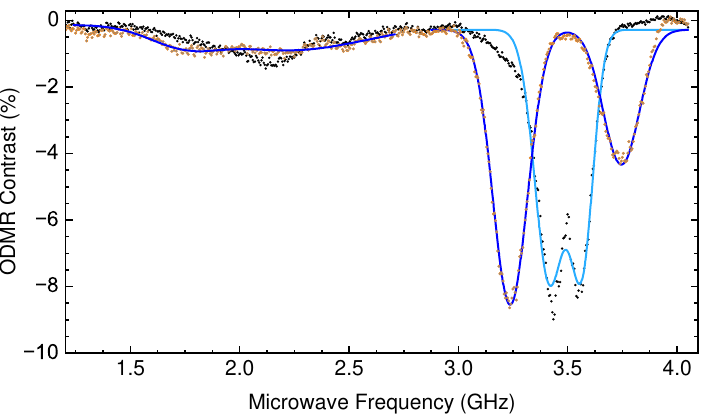}
\caption{\textbf{Fitting of the ODMR for extracting E and D parameters.} A gaussian fit with shared $D$, $E$ and $y_0$ has been used for the ground state transition (around $\SI{3.5}{\GHz}$) with $B=0$ and a bias magnetic field. With the magnetic field fixed the excited state transition with bias magnetic field (around $\SI{2}{\GHz}$) was fitted.}
    \label{fig:SI:ZFS-GS}
\end{figure*}
 
\subsubsection{Flake irradiation and transfer}
\label{sec:irradiation}
The flakes used for the DBR and ODMR measurements were exfoliated from commercially available hBN crystals (2D semiconductors). The flake was first transferred on a Si/SiO$_2$ substrate, and then irradiated with a \SI{30}{\kilo\electronvolt} nitrogen ion beam at a fluence of \SI{1e14}{ions\per\centi\meter\squared}. We then transferred the irradiated flakes onto the mesa sample using an HQ Graphene transfer stage. For the transfer, we used a PVC foil attached to a PDMS dome as the stamping layer. The hBN flake selected for the open-cavity measurements had a thickness of approximately 100~nm, as confirmed by AFM measurements shown in Supplementary Fig.~\ref{AFM}. 

Raman and photoluminescence (PL) spectra were acquired at room temperature using a Horiba LabRAM HR800 Raman system. The PL measurements were performed with a 600~lines~mm$^{-1}$ grating, while the Raman spectra were measured using an 1800~lines~mm$^{-1}$ grating. A 50$\times$ objective lens (Olympus, NA~=~0.6) was used to focus the excitation beam. During all measurements, the laser power was kept constant at approximately 1~mW.
The Raman spectrum of the sample, shown in Supplementary Fig.~\ref{Raman-Pl}(a), exhibits characteristic peaks at 1370.20~$\pm$~0.16~cm$^{-1}$ and 1299.62~$\pm$~0.93~cm$^{-1}$, assigned to the $E_{2g}$ and $D_1$ vibrational modes, respectively. The Si Raman peak at 520.7~cm$^{-1}$ was used as a reference for peak calibration. The measured Raman modes are in good agreement with previously reported values. The corresponding PL spectrum acquired from the same sample is shown in the right panel of  Supplementary Fig.~\ref{Raman-Pl}(a).
The Raman spectra were fitted using Lorentzian functions, while the PL spectrum was fitted using Gaussian functions. From these fits, the integrated areas of the $D_1$ peak, $A_{D1}$, the $E_{2g}$ peak, $A_{E_{2g}}$, and the PL emission, $A_{PL}$, were extracted. The ratio $(A_{D1}+A_{PL})/A_{E_{2g}}$ was found to be 2.42. Based on the analysis reported in Ref.\cite{Patra2025}, the spin-defect density was estimated to be approximately $10^{16}$~defects~cm$^{-3}$.
The spatial distribution of the defect-related emission was obtained from the PL map of the same flake, shown in Fig.~\ref{Raman-Pl}(b). The map was generated by integrating the PL intensity around 800~nm within a spectral window of $\pm 150$~nm, covering the broad near-infrared emission band associated with the $\mathrm{V_B^-}$ centres. The emission is observed across the flake area, indicating that the irradiation process generated optically active defect centres within the selected region.
The relatively uniform PL intensity is important for the cavity measurements, as it allows the emitter ensemble to be probed under comparable optical conditions. A region with homogeneous emission was therefore selected for the open-cavity experiments, marked by the white star in Fig.~\ref{Raman-Pl}(b), to minimise variations arising from local defect density, flake thickness, or spatial inhomogeneity. The same selected area was used to compare the uncoupled and cavity-coupled emission response and, separately, to estimate the density of boron-vacancy defects.

 \subsection{DBR deposition and cavity alignment}
 \label{sec:cavityalign}
 The bottom DBR mirror consists of six pairs of SiO$_2$/TiO$_2$ layers with thicknesses of approximately 136~nm and 90~nm, respectively. The top mirror comprises four pairs of the same dielectric layers of thickness 145 nm and 95 nm was deposited onto pre-patterned lenses on SiC substrate. Both mirrors were fabricated using dual-ion-beam sputter deposition (Nordiko 3000) at a base pressure of approximately $10^{-4}$~mbar.
 The alignment of the tunable open cavity and the angle-resolved measurements were performed in a custom-built optical setup. A 545~nm pulsed laser from a Coherent Mira 900 was aligned along the excitation path. A 50$\times$ long-working-distance objective with NA~=~0.65 was used for excitation and collection, while a $k$-space lens placed at the back focal plane enabled angle-resolved imaging. A slit was introduced after the $k$-space lens to spatially filter the emission before detection with an Andor spectrometer equipped with a Newton EMCCD camera.
For stable open-cavity operation, the bottom and top mirrors must be aligned as parallel as possible. We therefore followed a step-by-step alignment procedure, shown in Fig.~\ref{cavity_align}. First, the hBN flake on the bottom mirror was brought into focus with all lenses and the objective in the optical path, and this position was defined as the zero position of the closed-loop Attocube stage. The bottom mirror was then moved back, and the milled concave lens structure on the top mirror was brought into focus and also assigned a fixed reference position for the stages.
In the next step, all lenses and the objective were removed from the optical path, and the top mirror was moved away. The bottom mirror was returned to its reference position, and the reflected zero-order spot on the CCD was centred by adjusting the tilt of the bottom mirror. After this correction, the bottom mirror was moved back again, the top mirror was brought to its reference position, and the same angular correction was performed for the top mirror. This procedure ensured that both mirrors were aligned with respect to the optical axis before the cavity was closed.
After the angular correction, the lenses and objective were reinserted and realigned, starting from the focusing lens. The top mirror was fixed at the selected concave lens position, and the bottom mirror carrying the hBN flake was slowly approached towards it. At a certain mirror separation, the concave lens and the flake could be imaged together, as shown in Fig.~\ref{cavity_align}(d). The spatial slit was then positioned to collect emission only from the selected lens region, as shown in Fig.~\ref{cavity_align}(e). Finally, the laser spot was aligned to the centre of the concave lens. The corresponding image confirms that the excitation is incident on the selected lens and that the cavity supports Laguerre--Gaussian-like modes in real space, as shown in Fig.~\ref{cavity_align}(f).  

 \subsection{Angle resolved measurements}
 \label{sec:arm}
 Fig.~\ref{arm} shows the concept of the angle-resolved photoluminescence measurements; the emitted light was collected using a 50$\times$ objective (NA = 0.65), and its back focal plane was imaged onto the spectrometer through a Fourier-space lens system. The effective magnification of the Fourier imaging system was calculated as $M_{\mathrm{BFP}}=1.2$. 
 The effective objective focal length was estimated as $f_{\mathrm{obj}}=f_{\mathrm{tube}}/M_{\mathrm{obj}}$=4~mm,
 assuming a reference tube lens of 200~mm. The recorded CCD pixel positions were converted into emission angles using $\sin\theta=(i-i_0)p/(M_{\mathrm{BFP}}f_{\mathrm{obj}})$, where $i_0$ is the zero-angle pixel position and $p=16~\mu$m is the detector pixel size. The in-plane wave vector was subsequently calculated as $k_{\parallel}=(2\pi/\lambda)\sin\theta$, where $\lambda$ is the emission wavelength. The wavelength-dependent momentum coordinates were interpolated onto a uniform $k_{\parallel}$ grid, with the accessible angular range limited by the objective's numerical aperture.

\section{Cavity coupled PL }
\subsection{Estimation of airgap}
\label{sec:ag}
The air gap between the top and bottom mirrors was estimated by comparing the experimentally observed $k$ space image with transfer-matrix method (TMM) calculations in Fig.~\ref{air-gap}. In the $k$-space measurement, two cavity-mode families were observed at approximately 884~nm and 733~nm. The air-gap thickness in the TMM model was then varied until the calculated cavity modes matched these experimentally observed wavelengths. From this comparison, the air gap at the resonance condition between the cavity mode and the defect emission was estimated to be approximately 720~nm. This value was subsequently used as an input parameter for the FDTD simulations of the full-cavity configuration.

\subsection{Normalization factor}
\label{sec:NF}
We normalised the photoluminescence intensity by accounting for the effective excitation and collection area in the half-cavity and full-cavity configurations. In the half-cavity configuration, the excitation beam with an incident diameter of approximately 1~mm enters through a 50$\times$ objective with NA~=~0.65 and a tube lens focal length of 200~mm. Using the corresponding spot-size conversion, the effective excitation diameter at the sample plane was estimated to be approximately 5~$\mu$m.

For the full-cavity configuration, the relevant optical area was estimated from the fundamental cavity-mode waist. The beam waist radius was calculated using
\begin{equation}
w_0 =
\left[
\frac{\lambda_0}{\pi}
\sqrt{L(R-L)}
\right]^{1/2},
\end{equation}
where $\lambda_0$ is the resonance wavelength, $L$ is the effective cavity length, and $R$ is the radius of curvature of the concave lens. The radius of curvature was obtained from the measured lens geometry using
\begin{equation}
R = \frac{a^2 + d^2}{2d},
\end{equation}
where $a$ is the radius of the concave lens and $d$ is its depth\cite{Hunger2010,Kogelnik1966LaserBA}.
The effective mode area in the full-cavity and half-cavity configuration was then calculated by using the waist radius and the radius of the spot in the half cavity. The photoluminescence spectra were normalized by the corresponding effective area to compare the emission intensity in units of counts~s$^{-1}$~area$^{-1}$. This area normalization allows a reliable comparison between the half-cavity reference and the cavity-coupled emission, where the optical mode is confined to a smaller spatial region shown in the Figure 3(d) in main text and Figure~\ref{PL raw}.

\section{Lifetime fitting}
 \label{sec:IRF}
The TRPL decay curves were analysed by reconvolution fitting using a biexponential decay model and a Gaussian instrument response function (IRF)\cite{Trinh2021}. The independently measured IRF was fitted with a Gaussian peak and a constant background, yielding a standard deviation of $\sigma_{\mathrm{IRF}}\approx 35 $~ps. This corresponds to a full width at half maximum of
\begin{equation}
\mathrm{FWHM}_{\mathrm{IRF}}
=2\sqrt{2\ln 2}\,\sigma_{\mathrm{IRF}}
\approx82~\mathrm{ps}.
\end{equation}

For the convolution, the background-subtracted IRF was represented by a normalized Gaussian,
\begin{equation}
g(t)=\frac{1}{\sqrt{2\pi}\sigma_{\mathrm{IRF}}}
\exp\left(-\frac{t^2}{2\sigma_{\mathrm{IRF}}^2}\right).
\end{equation}
The underlying emission decay was described by
\begin{equation}
D(t)=\Theta(t)\sum_{i=1}^{2}A_i e^{-t/\tau_i},
\end{equation}
where $\Theta(t)$ is the Heaviside step function, $A_i$ are the decay amplitudes, and $\tau_1$ and $\tau_2$ denote the fast and slow decay times, respectively.

The measured transient was modelled as
\begin{equation}
I_{\mathrm{fit}}(t)
=B+\int_0^\infty g(t-t_0-u)D(u)\,\mathrm{d}u,
\end{equation}
where $t_0$ accounts for the temporal offset and $B$ is a constant background. The convolution was evaluated analytically, giving
\begin{equation}
I_{\mathrm{fit}}(t)=B+\sum_{i=1}^{2}I_i(t),
\end{equation}

\begin{equation}
I_i(t)=\frac{A_i}{2}
\exp\left[
\frac{\sigma_{\mathrm{IRF}}^2}{2\tau_i^2}
-\frac{t-t_0}{\tau_i}
\right]
\mathrm{erfc}\left[
\frac{\sigma_{\mathrm{IRF}}^2/\tau_i-(t-t_0)}
{\sqrt{2}\sigma_{\mathrm{IRF}}}
\right],
\end{equation}
where $\mathrm{erfc}$ denotes the complementary error function. The convolved model was fitted to the experimental data using nonlinear least squares. The Gaussian standard deviation $\sigma_{\mathrm{IRF}}$, rather than its FWHM, enters the fitting function directly. This procedure accounts for instrumental temporal broadening when extracting the PL decay times.

\section{ODMR}
\label{sec:ODMR}
The ODMR was measured in a home-built confocal microscope. We use a \SI{473}{\nm} Cobolt 06-MLD laser for excitation that is focused through a Mitutoyo Apo NIR HR x100 objective lens onto the sample. The PL is collected by the same objective lens and separated from the excitation by a dichroic mirror. It is subsequently collected with an Excelitas SPCM-AQRH-14 single photon detector. The incoming counting events are recorded by a TimeTagger20 time-correlated single-photon counting module by Swabian Instruments.
Microwaves were generated by a signal generator SG380 from Stanford Research Systems and modulated by a microwave switch. Additionally, the microwave power was amplified by a 25W VBA-0660 power amplifier from Vectawave.
Microwave magnetic fields are delivered by a coplanar waveguide (CPW) on silicon (see \autoref{fig:SI:CPW}). The silicon wafer is covered in a \SI{280}{\nm} thick SiO$_2$ and subsequently a thin Cr (\SI{60}{\nm}) and Au (\SI{150}{\nm}) layer is deposited. 
Afterwards, an nitrogen-irradiated hBN flake is transferred onto the center conductor of the stripline (see \autoref{fig:SI:Flake})
Measurements were conducted by boxcar averaging combined with a fast sweep over the microwave frequency.
A spin triplet experiences a zero field splitting of the form
\begin{equation}
    H_\mathrm{ZFS}= h D\left( S_z^2-\frac{S(S+1)}{3}\right) + h E(S_x^2-S_y^2) \notag
\end{equation}
where $D$ and $E$ are derived from the zero-field tensor $\mathbf{\hat{D}}$ with the relation $D=3/2 D_{z}$ $E=1/2(D_{x}-D_{y})$ and $D_i$ being the component in the direction of the axis $i$ of the primary axis system of $\mathbf{\hat{D}}$.
Under perfect symmetry only the value of $D$ is non-zero. 
The energy eigenstates take the values $h\nu_\mathrm{+1} = h(D+E)$ and $h\nu_\mathrm{-1}=h(D-E)$. With the application of an external magnetic field, these values become $h\nu_\mathrm{+1} = h\left(D+\sqrt{E^2+g\frac{\mu_\mathrm{B}}{h}B}\right)$ and $h \nu_\mathrm{-1}=\left(D-\sqrt{E^2+g\frac{\mu_\mathrm{B}}{h}B}\right)$.
The ZFS parameters are measured by the ODMR spectrum shown in \autoref{fig:SI:ZFS-GS} under zero field (no external magnetic field) and with a bias (external magnetic field). Two Gaussian curves with mean value $\mu_\mathrm{+1} = \nu_\mathrm{+1}$ and $\mu_\mathrm{-1}= \nu_\mathrm{-1}$ have been fitted to the data.
The fit was carried out for the GS transition using a global fit between the zero field and the biased measurements. Both $D$ and $E$ values have been shared for both fits and $B$ was fixed at 0 for the zero field measurement. 
The calculations yield $D$, $E$ and $B$ as described in the main text.
The ES transition at $B=0$ shows significant deviation from two gaussians, rendering the fitting prohibitively difficult. When fitted with the $B$ value from the GS fit, we can only approximate $E_\mathrm{ES}=0$.

% Example supplementary figure. Uncomment after adding the image file.
% Store the file inside the Supplementary_Figures folder.
% \begin{figure}[t]
%   \centering
%   \includegraphics[width=\textwidth]{Figure_S1.png}
%   \caption{Caption for the first supplementary figure.}
%   \label{fig:S1}
% \end{figure}

% Example supplementary table.
% \begin{table}[t]
%   \centering
%   \caption{Caption for the first supplementary table.}
%   \label{tab:S1}
%   \begin{tabular}{cc}
%     \hline
%     Parameter & Value \\
%     \hline
%     Example & 1 \\
%     \hline
%   \end{tabular}
% \end{table}

%\printbibliography[heading=bibintoc,title={Supplementary References}]

\clearpage
\printbibliography[heading=bibintoc,title={Supplementary References}]

\end{refsection}
\end{document}

\fi

\end{document}